\documentclass[aps,a4paper,showpacs,twocolumn]{revtex4}
\usepackage{graphicx}
\usepackage{amsmath}
\usepackage{amssymb}
\usepackage{enumerate, color}
\usepackage{subfigure}
\usepackage{tabularx}
\usepackage{epstopdf}
\usepackage{lipsum}
\usepackage{color}
\newcommand{\be}{\begin{equation}}
\newcommand{\ee}{\end{equation}}
\newcommand{\ben}{\begin{eqnarray}}
\newcommand{\een}{\end{eqnarray}}
\newcommand{\bes}{\begin{subequations}}
\newcommand{\ees}{\end{subequations}}

\newcommand{\bb}{\bibitem}

\begin{document}
\title{New models for asymmetric kinks and branes}
\author{ R. Menezes$^{1,2}$}
\author{D.C. Moreira$^3$}
\affiliation{$^1$Departamento de F\'isica, Universidade Federal da Para\'iba, 58051-970, Jo\~ao Pessoa, PB, Brazil}
\affiliation{$^2$Departamento de Ci\^encias Exatas, Universidade Federal da Para\'\i ba, 58297-000 Rio Tinto, PB, Brazil}
\affiliation{$^3$Departamento de F\'\i sica, Universidade Federal de Campina Grande, 58109-970 Campina Grande, PB, Brazil}
\begin{abstract}
We investigate new models for scalar fields in flat and curved spacetime. We note that the global reflection symmetry of the potential that identify the scalar field model does not exclude the presence of internal asymmetries that give rise to asymmetric structures. Despite the asymmetry, the new structures are linearly stable and in the braneworld scenario with an extra dimension of infinite extend, they may generate new families of asymmetric thick branes that are robust against small fluctuations in the warped geometry.   
\end{abstract}
\maketitle

%%%%%%%%%%%%%%%%%%%%%%%%%%%%%%%%%%%%%%%%%%%%%
%%%%%%%%%%%%%%%%%%%%%%%%%%%%%%%%%%%%%%%%%%%%%
\section{Introduction} 
%%%%%%%%%%%%%%%%%%%%%%%%%%%%%%%%%%%%%%%%%%%%%
%%%%%%%%%%%%%%%%%%%%%%%%%%%%%%%%%%%%%%%%%%%%%

The study of scalar fields has been very productive in presenting relevant nonlinear models and their defect structures. The subject is of direct interest to high energy physics \cite{rajaraman,v,sut}, and the presence of solutions of topological or non-topological nature may play important role in several areas of Physics, as one sees, for instance, in Refs.~\cite{ap1,ap2,ap3,ap4,ap5,ap6,ap7}. 

When the model is defined in $(1,1)$ space-time dimensions, described by a single real scalar field and governed by a potential with a set of degenerate minima, the field solution linking two consecutive minima are {\it kinks}, which are defect structures of topological nature. In general, the potential can be polynomial or non-polynomial; in the case of a polynomial potential, the set of minima is finite, and one has a finite number of topological sectors with topological or kinklike solutions. When the potential is non-polynomial, however, thare are other possibilities. For example, if the potential is periodic as in the sine-Gordon model \cite{rubinstein,caudrey} and in many of its generalizations and deformations \cite{ba1,blm}, there may be an infinite set of degenerate minima, and families of non-equivalent topological sectors may appear. 

Beyond the $\phi^4$, $\phi^6$, and the sine-Gordon models, there are other possibilities which we will study in this work. Specifically, we will focus attention on non-polynomial interactions, governed by potentials that engender the $\mathbb{Z}_2$-symmetry and  have a finite number of topological sectors, controlled by the parameters that identify the potential. We have found two distinct classes of models of this kind, described by the asymptotic behavior of the potential: in one class of models, the potential diverges as the field increases toward higher and higher values; in the other class, the potential reaches a constant value asymptotically.

In these two families of models, one finds topological structures with energy densities that are asymmetric. As one knows, finding asymmetric solutions is important due to its relation to thick branes, since it is a well know fact that scalar fields can induce braneworld scenarios where the scalar field acts as a source of gravity \cite{RS,gw,fre,csaki}. In many of those models, the field solution leads to reflection invariance, thus generating symmetric branes; see, e.g., Refs.~\cite{bm, bfg,dutra2,liu1,liu2,blmm,gremm,almeida,brito}. However, there are some asymmetric solutions, as found, for instance, in \cite{melfo2,melfo1,dutra1,bmr,bmm1,lobo}. In different backgrounds, investigations of asymmetric branes point that a brane in between two distinct five dimensional spacetimes can induce cosmic acceleration on the brane with late-time de Sitter solutions and modifications of gravity in the infra-red limit \cite{padilla1,  kraus,padilla2,gregory,toporensky}. However, there are not many works on cosmology of thick branes, even in the symmetric case \cite{ahmed}. Thus, searching for new analytically solvable asymmetric thick brane models is interesting to expand the possibility of exploring new issues concerning the cosmic evolution.
 
The paper is organized as follows. In Sec.~II one makes a review of BPS solutions and its stability, specifying some fundamental aspects concerning topological solutions in Classical Field Theory. Furthermore, it is made a discussion about the deformation procedure and how it leads to new models. After that, we add a review on the main results presented in \cite{bllm}. The new models are presented in Sec.~III, where we found  topological solutions and make its stability analysis. We also discuss a new possibility to classical trajectories of the mechanical analog of Model 2. In Sec.~IV it is made an analysis of the thick brane configuration derived from the new models.

\section{Generalities}
%%%%%%%%%%%%%%%%%%%%%%%%%%%%%%%%%%%%%%%%%%%%%
%%%%%%%%%%%%%%%%%%%%%%%%%%%%%%%%%%%%%%%%%%%%%
\subsubsection{BPS solutions in Classical Field Theory}
%%%%%%%%%%%%%%%%%%%%%%%%%%%%%%%%%%%%%%%%%%%%%
%%%%%%%%%%%%%%%%%%%%%%%%%%%%%%%%%%%%%%%%%%%%%
The simplest case where one can deal with the dynamics of a scalar field  is written in terms of a Lagrangian in the form \cite{rajaraman,v}
\be\label{ori}
\mathcal{L}(\phi,\partial_{\mu}\phi)=\frac{1}{2}\partial_{\mu}\phi\partial^{\mu}\phi-V(\phi).
\ee
In this case, $\phi=\phi(x,t)$ is a scalar field in a $(1,1)$ Minkowski spacetime. The metric tensor is 
$\eta_{\mu\nu}={\rm diag}\,(1,-1)$ and the quantity $V(\phi)$ is the potential of the model. The equation of motion for the scalar field is 
given by the Euler-Lagrange equation, which has the form
\begin{equation}\label{equationofmotion}
\partial_{\mu}\partial^{\mu}\phi+\frac{dV}{d\phi}=0.
\end{equation}
We can reduce our problem by studying static configurations, with the time evolution being reconstructed through a Lorentz boost. In this approach we only have to deal with a second order differential equation, which is given by
\begin{equation}\label{2ode}
\frac{d^2\phi}{dx^2}=\frac{dV}{d\phi}
\end{equation}
plus some boundary conditions. Note that, in general, the form of the potential $V(\phi)$ may lead to a highly nonlinear differential equation, which can make the analysis of (\ref{2ode}) very hard, or even impossible to be accomplished analytically. Thus it is important to know how to implement methods that simplify the description of the systems represented by (\ref{2ode}), to find analytical solutions. 

The energy-momentum tensor associated to the Lagrangian (\ref{ori}) is $T^{\mu\nu}=\partial^{\mu}\phi\partial^{\nu}\phi-\eta^{\mu\nu}\mathcal{L}$
 and the energy density $\rho (x)$  is just the $T^{00}$  component, which can be written as
\bes\label{rho1}\ben
\rho (x)&=&\frac{1}{2}\phi'^2+V(\phi)\\
&=&\frac{1}{2}\left(\phi'\mp\sqrt{2V(\phi)}\right)\pm\phi'\sqrt{2V(\phi)}.
\een\ees 

The energy density (\ref{rho1}) gives a clue about the  way one can proceed to find useful approaches in the treatment of models described by (\ref{ori}): by writting $V(\phi)$  as a square of some function, one can eliminates the square root present in this formula.  We choose a function $W(\phi)$ such that
\begin{equation}\label{potv}
V(\phi)=\frac{1}{2}\left(\frac{dW}{d\phi}\right)^2.
\end{equation}
This simple choice makes the analysis of the system easier, in general, because now we have to deal with first order differential equations
\begin{equation}\label{1order}
\frac{d\phi}{dx} =\pm W_{\phi},
\end{equation}
where $W_{\phi}={dW}/{d\phi}$. A particular and important feature due to the presence of the {\it first order formalism} in this type of systems is that the energy density becomes
\begin{equation}
\rho (x)=\frac{1}{2}\left(\phi'\pm W_{\phi}\right)\pm\frac{dW}{dx}.
\end{equation}
It is such that for solutions of the first order equations (\ref{1order}), the energy is minimized to
\begin{equation}\label{Ebps}
E_{BPS}=|W\left(\phi (\infty)\right)\!-\!W\left(\phi \left(-\infty\right)\right)|. 
\end{equation}
$E_{BPS}$ is called {\it BPS energy} \cite{bps}.

When the potential we are dealing has a set of degenerate global minima, it necessarily has topological sectors supporting {\it kink-like} solutions. Each kink is characterized by a topological current usually written as
\begin{equation}\label{current}
j^{\mu}=\epsilon^{\mu\nu}\partial_{\nu}\phi,
\end{equation} 
where $\epsilon^{\mu\nu}$ is the antisymmetric symbol in two dimensions with $\epsilon^{01}=1$. The current (\ref{current}) has an associated conserved quantity, denoted here by $Q$, given by
\begin{equation}
Q=\phi (\infty)-\phi (-\infty)
\end{equation}
wich is called {\it topological charge} of the kink.

One can investigate the linear stability of BPS solutions by making a small time-dependent perturbation around the static solutions, which can have the form  $\phi(x,t)=\phi(x)+\sum_n \eta_n(x)\cos (\omega_n t)$, for small perturbation, and then plugging $\phi(x,t)$ in (\ref{equationofmotion}).  In this case, one gets the Schr\"odinger-like  equation
\begin{equation}\label{stabilityequation}
\left(-\frac{d^2}{dx^2}+v(x)\right)\eta_n(x)=\omega_n^2\eta_n(x)
\end{equation}
where the stability potential is
\bes\label{stabilitypotential}\ben
v(x)&=&\frac{d^2V}{d\phi^2}\biggr{|}_{\phi=\phi(x)}\\
&=&W_{\phi\phi}\bigr{|}_{\phi=\phi(x)}+W_{\phi\phi\phi}W_\phi\bigr{|}_{\phi=\phi(x)}
\een\ees
We know that the operator ${H}=-d^2/dx^2+v(x)$, due to the relation (12), can be written as ${H}=S^{\dag}S$ with $S^{\dag}=- d/dx-W_{\phi\phi}$. So, it is non negative and we have $\omega_n^2\geq0$ for all values of $n$. It ensures the stability of the system, since there is no negative energy modes.  Moreover, we have at least one bound state, which is associated with the translational invariance of our solutions, which is given by $\eta_0(x)=d\phi/dx.$ It is the zero mode.
%%%%%%%%%%%%%%%%%%%%%%%%%%%%%%%%%%%%%%
\subsubsection{The Deformation Procedure}
%%%%%%%%%%%%%%%%%%%%%%%%%%%%%%%%%%%%%%
The Deformation Procedure consists in a way for generating new models in Field Theory with non trivial dynamics derived from models wich already has  a well known behavior \cite{blm}. The key point in this method is that the new model obtained from the deformation procedure keeps the main general aspects of the previous theory. In particular, if we start working with a theory that has BPS solutions with a first order formalism, the new model obtained after the deformation will also have.

The method basically consists in, given a theory with a Lagrangian of the form (\ref{ori}), a change in the field is made through a {\it deformation function} $f$, which is a function of the other field $\chi$, such that
\begin{equation}\label{transf1}
\phi\longrightarrow f\left(\chi\right).
\end{equation}
As a consequence, now the new field $\chi$ is governed by the Lagrangian
\begin{equation}\label{defmod}
\bar{\mathcal{L}}(\chi,\partial_{\mu}\chi)=\frac{1}{2}\partial_{\mu}\chi\partial^{\mu}\chi-U(\chi),
\end{equation}
where a new potential (a new theory) is defined by $U(\chi)=V\left(\phi\rightarrow f(\chi)\right)/f_{\chi}^2.$
Assuming that  $V\left(\phi\right)$ can be written as (\ref{potv}), i.e.  as a square of some function, so the same is valid for $U\left(\chi\right)$. Thus, we are able to find the first order differential equation for the new field
\begin{equation}\label{eq1}
\frac{d\chi}{dx}= \pm \bar{W}_{\chi},
\end{equation}
with $\bar{W}_\chi=W_\phi(\phi\to f(\chi))/f_\chi.$

The Deformation Procedure is particularly simple when one looks for the solutions of the new models. Since the solution for the field of the starting model is already known, the form of the solution of the new model can be directly obtained from (\ref{transf1}). The solution, in this case, is just the inverse of the deformation function $f$ applied to the field solution $\phi(x)$ of the starting model. That is to say
\begin{equation}\label{sol}
\chi(x)=f^{-1}(\phi(x)).
\end{equation}
Thus, in fact, when we take an appropriate deformation function for (\ref{transf1}), basically we are automatically providing the solution of the new model. The inherent first order formalism present in this procedure, here represented by (\ref{eq1}), implies that there exists a new $\bar{W}$-function defined for the model and that the BPS energy of the new model is given by
\begin{equation}
\bar{E}_{BPS}=|\bar{W}\left(\chi (\infty)\right)-\bar{W}\left(\chi(-\infty)\right)|.
\end{equation}
where $\bar{W}\left(\chi\right)$ is the $W$-function defined in the new model.

\subsubsection{Previous work}

Previously, in \cite{bllm}, the Deformation Procedure applied on the $\phi^4$ model through the function
\begin{equation}
f(\chi)=\cos \left[a\left(\cos^{-1} \chi -\frac{m}{a}\pi\right)\right]
\end{equation}
provided us with new models, represented by the potentials
\bes\label{previous}\ben
&&\bar{u}_s^a\left(\chi\right)=\frac{\left(1-\chi^2\right)^2}{2a^2} \mathcal{U}_{a-1}^2\left(\chi\right),~\text{if}~m=0,\pm 1,\pm 2,...\;\;\;\;\;\;\;\;\;
\\
&&\bar{u}_c^a\left(\chi\right)=\frac{\left(1-\chi^2\right)}{2a^2}\mathcal{T}_a^2\left(\chi\right),~\text{if}~m=\pm 1/2,\pm 3/2,...\;\;\;\;\;\;\;\;\;
\een\ees
The models above are expressed in terms of a class of special functions, which are the Chebyshev Polynomials. Such polynomials can be represented  in terms of elementary trigonometric functions by,
\bes\ben
&&\mathcal{U}_a\left(\chi\right)=\frac{\sin \left((a+1)\cos^{-1} \chi\right)}{\sin \cos^{-1}\chi}
\\
&&\mathcal{T}_a\left(\chi\right)=\cos\left(a\cos^{-1}\chi\right).
\een\ees
$\mathcal{T}_a\left(\chi\right)$ and $\mathcal{U}_a\left(\chi\right)$ are called Chebyshev Polynomials of First and Second kind, respectively. An interesting feature of these functions is that its obey the {\it Pell Equation},  $\mathcal{T}_a^2\left(\chi\right)-\left(\chi^2-1\right)\mathcal{U}_{a-1}^2\left(\chi\right)=1$,
which is an interesting Diophantine equation, well-known in Number Theory. In these models, the $a$-parameter is used to generate and control the number of the new emergent topological sectors while the $m$-parameter is a phase acting to separate each one, so the field solution changes at each topological sector and is related with both $a$ and $m$. The general form of the field solution is
\begin{equation}
\chi(x)=\cos \left[\frac{\cos^{-1}\left(\pm\tanh x\right)+m\pi}{a}\right].
\end{equation}
The plus signal is related with the $\phi^4-kink$ and the minus signal is related with the $\phi^4-antikink$. The study of neutral bound states in the quantum version of this model was presented  in \cite{mussardo}.

%%%%%%%%%%%%%%%%%%%%%%%%%%%%%%%%%%%%%%%%%%%%%%%%%%%%%
%%%%%%%%%%%%%%%%%%%%%%%%%%%%%%%%%%%%%%%%%%%%%%%%%%%%%
\section{Models}
%%%%%%%%%%%%%%%%%%%%%%%%%%%%%%%%%%%%%%%%%%%%%%%%%%%%%
%%%%%%%%%%%%%%%%%%%%%%%%%%%%%%%%%%%%%%%%%%%%%%%%%%%%%
The purpose of this section is to take a step forward and  look at what new properties can be obtained by virtue of the  previous models (\ref{previous}). We are interested in the model generated by the deformation function
\begin{equation}\label{d1}
g(\theta)=\tanh\theta.
\end{equation}
From the Deformation mechanism, the new models we are interested in are 
\bes\label{potentials}\ben
&&u_s^a\left(\theta\right)=\frac{1}{2a^2}\mathcal{U}_{a-1}^2\left(\tanh\theta\right),
\\
&&u_c^a\left(\theta\right)=\frac{1}{2a^2}\cosh ^2\theta~\mathcal{T}_a^2\left(\tanh\theta\right),
\een\ees
for $m=0,\pm 1,\pm 2,...$ and for $m=\pm 1/2,\pm 3/2,...$, respectively. These models still are in terms of the Chebyshev Polynomials and it is interesting to note that, despite the freedom in the possible values assumed by $m$ {\it a priori} in the previous model, it aways lead us to a finite number of topological sectors, since $a$ is finite in each case. The quadratic form of the  potentials of these models implies that there is a $W$-function associated with this model, but unfortunately it is not possible to find its general analytical form. However, it is always possible to find it for a chosen $a$. The general solutions for the field $\theta(x)$ is obtained  by using the  Deformation Method, specifically the relation (\ref{sol}). It has to obey
\begin{equation}\label{gensolu}
\theta_m^a (x)=\tanh^{-1} \left[\cos\left( \frac{\cos^{-1}\left(\pm\tanh x\right)+m\pi}{a}\right)\right].
\end{equation}
For a given $a$, $m$ and $m\pm2ka$ represents degenerate solutions for any $k\in \mathbb{Z}$. The solution  (\ref{gensolu}) works for both potentials (\ref{potentials}), but the boundary conditions on the topological sectors implies some quantitative differences between the systems. It happens due to the fact that there must exists a compatibility relation between the minima of the potentials and the asymptotic behavior of each field solution, since they have to equal one another.

%%%%%%%%%%%%%%%%%%%%%%%%%%%%%%%%%%%%%%%%%%%%%%%
%%%%%%%%%%%%%%%%%%%%%%%%%%%%%%%%%%%%%%%%%%%%%%%
\subsection{Model 1}
%%%%%%%%%%%%%%%%%%%%%%%%%%%%%%%%%%%%%%%%%%%%%%%
%%%%%%%%%%%%%%%%%%%%%%%%%%%%%%%%%%%%%%%%%%%%%%%

We start the study of  the models (\ref{potentials}) looking at the case where $m=\pm 1/2,\pm 3/2,..$,  i.e., the model we are interested here is
\begin{equation}\label{potc}
u_c^a\left(\theta\right)=\frac{1}{2a^2}\cosh ^2\theta~\mathcal{T}_a^2\left(\tanh\theta\right).
\end{equation}
The function above represents a class of models ordered by the $a$-parameter. The first potentials contempled by (\ref{potc}) are
\begin{eqnarray}
\nonumber u_c^2\left(\theta\right)&=&\frac{1}{8} \cosh ^2\theta  \left(2 \tanh ^2\theta -1\right)^2\\
\nonumber u_c^3\left(\theta\right)&=&\frac{1}{18} \cosh ^2\theta  \left(4 \tanh ^3\theta -3 \tanh \theta \right)^2\\
\nonumber u_c^4\left(\theta\right)&=&\frac{1}{32} \cosh ^2\theta  \left(8 \tanh ^4\theta -8 \tanh ^2\theta +1\right)^2\\
\nonumber u_c^5\left(\theta\right)&=&\frac{1}{50} \cosh ^2\theta \left(16 \tanh ^5\theta-20 \tanh ^3\theta+ 5 \tanh \theta\right)^2
\end{eqnarray}
We can observe that in general these models appears as polynomials of $\tanh \theta$, wich reflects the highly non-linear profile of the models that we are dealing. Tha  shape of the first models of the class (\ref{potc}) are  depicted in FIG.\ref{fig1.1} and FIG.\ref{fig1.2} for odd and even values of $a$, respectively.
%%%%%%%%%%%%%%%%%%%%%%%%%%%%%%%%%%%%%%%%%%%%%%
%%%%%%%%%%%%%%%%%%%%%%%%%%%%%%%%%%%%%%%%%%%%%%
\setcounter{figure}{0}
\begin{figure}[ht]
\renewcommand{\thefigure}{1.1}
\centerline{\includegraphics[height=15em]{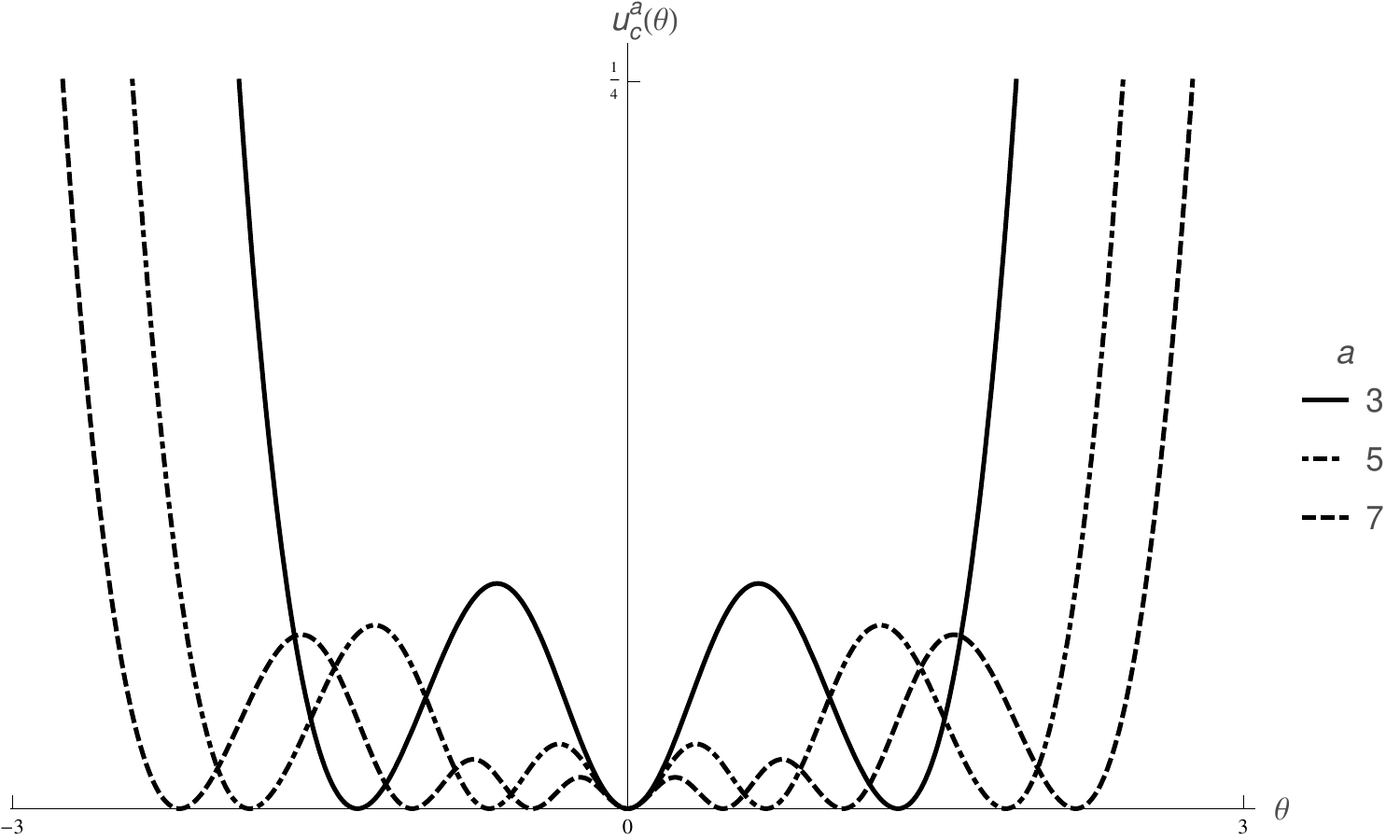}}
\caption{Potential $u_c^a\left(\theta\right)$ for some values of odd $a$.}\label{fig1.1}
\end{figure}
%%%%%%%%%%%%%%%%%%%%%%%%%%%%%%%%%%%%%%%%%%%%%%
%%%%%%%%%%%%%%%%%%%%%%%%%%%%%%%%%%%%%%%%%%%%%%
%%%%%%%%%%%%%%%%%%%%%%%%%%%%%%%%%%%%%%%%%%%%%%
%%%%%%%%%%%%%%%%%%%%%%%%%%%%%%%%%%%%%%%%%%%%%%
\begin{figure}[t]
\renewcommand{\thefigure}{1.2}
\centerline{\includegraphics[height=15em]{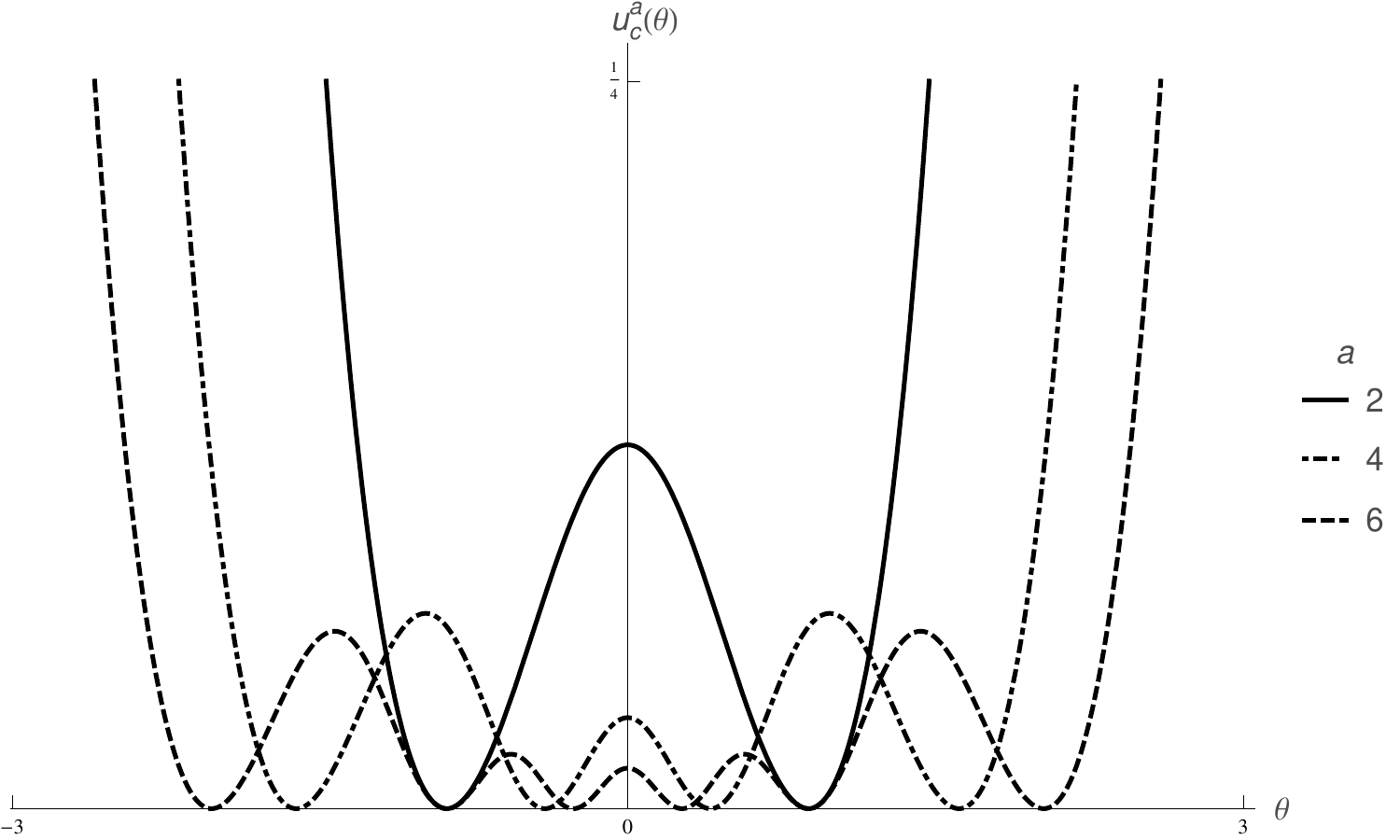}}
\caption{Potential $u_c^a\left(\theta\right)$ for some values of even $a$.}\label{fig1.2}
\end{figure}
%%%%%%%%%%%%%%%%%%%%%%%%%%%%%%%%%%%%%%%%%%%%%%%%

These models have a finite set of degenerate minima depending of each value of $a$, and the  $\cosh^2$ term imposes a divergent  exponential  behavior for the potentials as $\theta\rightarrow\pm\infty$. The first $W$-functions for these models are
\begin{eqnarray}\label{wfunctionc}
W_c^2(\theta)&=&\frac{1}{2} \left(\sinh\theta-4 \tan ^{-1}\left(\tanh \left(\theta/2\right)\right)\right)+c_2\\
\nonumber W_c^3(\theta)&=&\frac{1}{3} (\cosh \theta+4 \text{sech}\theta)+c_3\\
\nonumber W_c^4(\theta)&=&\sinh\theta\left(\frac{1}{4} +\text{sech}^2\theta \right)-2 \tan^{-1}\left(\tanh \theta/2\right)+c_4\\
\nonumber W_c^5(\theta)&=&\frac{1}{15} \left(3 \cosh^4\theta +18 \cosh 2\theta+2\right) \text{sech}^3\theta +c_5
\end{eqnarray}
where the $c_i$-numbers ($i=2,3, ...$) are constants wich does not play any role in the field content of the model, but have relevant implications when studying  asymmetric braneworld scenarios, as discussed in \cite{bmr}. 
The zeros of this  class of models are distributed  in the points  $\theta_{c, n}=\tanh^{-1} \left(\cos\left(\frac{2n-1}{2}\frac{\pi}{a}\right)\right)$, for $1\leq n\leq a$ and in the center one has
$$u_c^a(0)=\left\{
\begin{array}{cc}
0&\text{, if $a$ is odd}\\
\frac{1}{2a^2}&\text{, if $a$ is even.} \\
\end{array}
\right.$$
So, for odd $a$-values one has no topological sector passing through the point $\theta=0$, but for even $a$ it does.  It is also clear that we have $a/2$ different topological sectors for even $a$-values, and $(a-1)/2$ different topological sectors for odd $a$-values. The other sectors can be obtained when analyzing the $Z_2$-symmetry of the model.

%%%%%%%%%%%%%%%%%%%%%%%%%%%%%%%%%%%%%%%%%%%%%%
%%%%%%%%%%%%%%%%%%%%%%%%%%%%%%%%%%%%%%%%%%%%%%
\begin{figure}[t]
\renewcommand{\thefigure}{2}
\centerline{\includegraphics[height=15em]{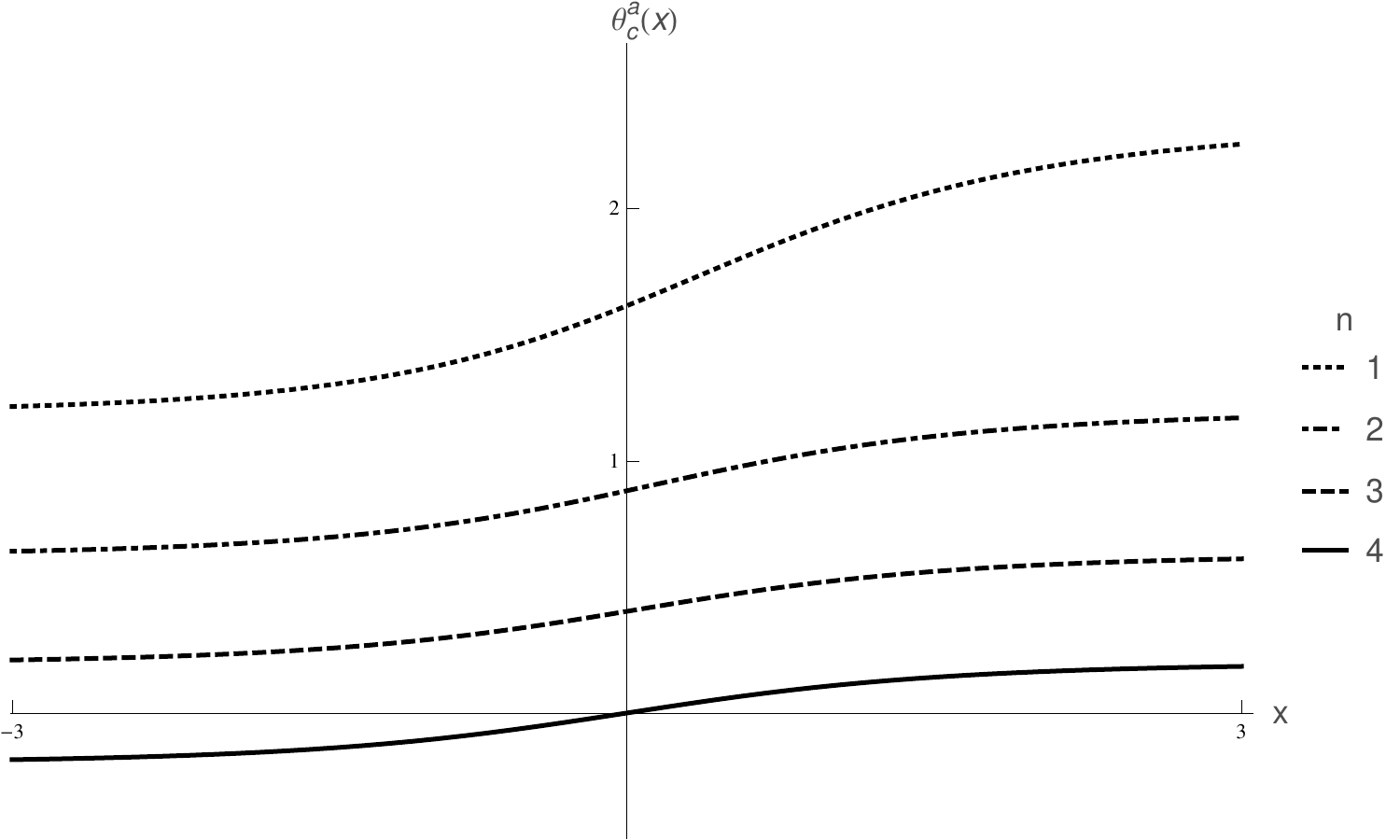}}
\caption{The field solution associated with the potential (\ref{potc}) for $a=8$ and $n=1,2,3$ and 4.}\label{fig2}
\end{figure}
%%%%%%%%%%%%%%%%%%%%%%%%%%%%%%%%%%%%%%%%%%%%%%
%%%%%%%%%%%%%%%%%%%%%%%%%%%%%%%%%%%%%%%%%%%%%%
%%%%%%%%%%%%%%%%%%%%%%%%%%%%%%%%%%%%%%%%%%%%%%
%%%%%%%%%%%%%%%%%%%%%%%%%%%%%%%%%%%%%%%%%%%%%%
\begin{figure}[t]
\renewcommand{\thefigure}{3}
\centerline{\includegraphics[height=15em]{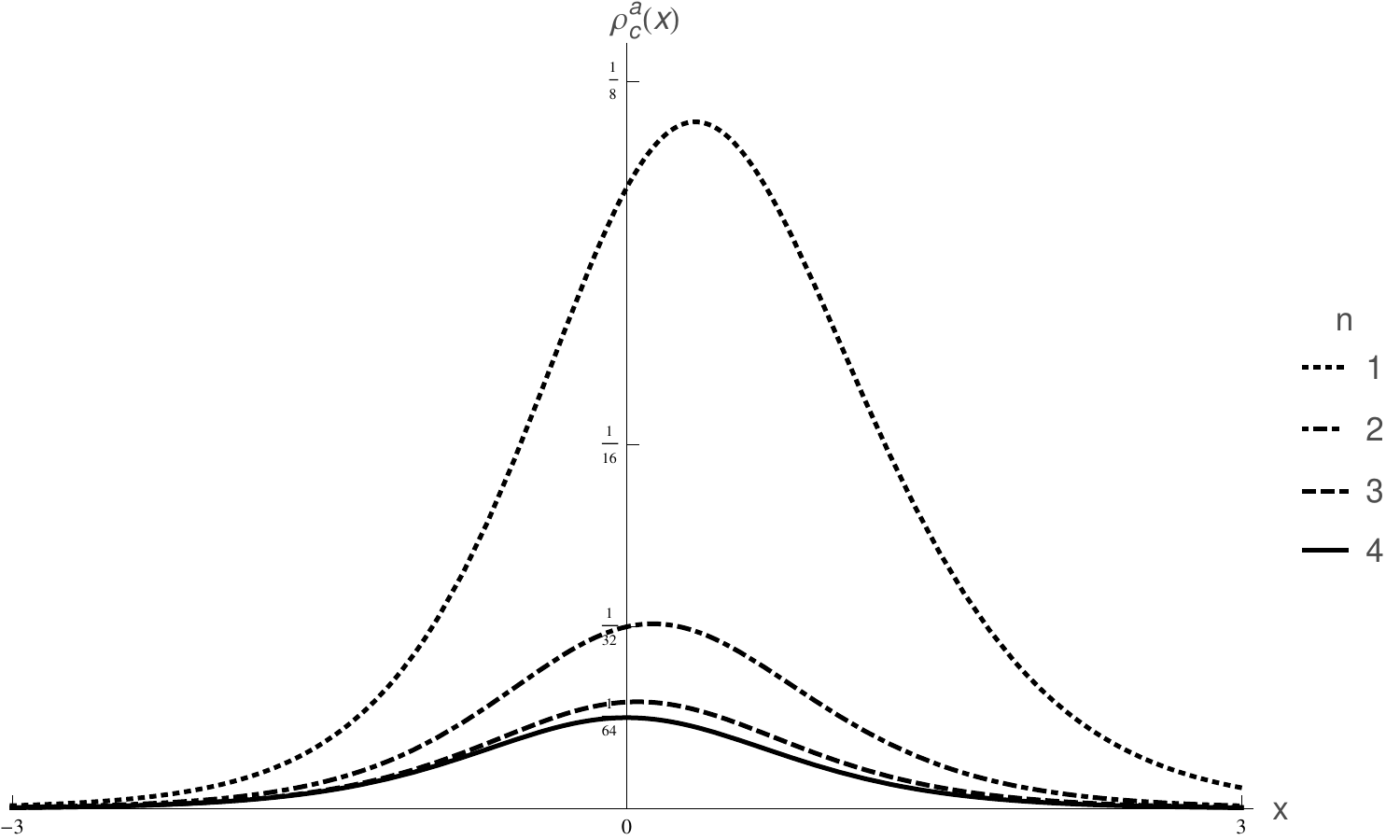}}
\caption{The energy density (\ref{edc}) for $a=8$ and $n=1,2,3$ and 4.}\label{fig3}
\end{figure}
%%%%%%%%%%%%%%%%%%%%%%%%%%%%%%%%%%%%%%%%%%%%%%
%%%%%%%%%%%%%%%%%%%%%%%%%%%%%%%%%%%%%%%%%%%%%%
The equation for the field in the model we are studying is the first order equation given by
\begin{equation}\label{1ordert}
\theta'=\frac{1}{a}\cosh \left(\theta\right)\mathcal{T}_a\left(\tanh \theta\right).
\end{equation}
Despite the complicated form, we already know that the solution of the equation (\ref{1ordert}) is given by (\ref{gensolu}) through the Deformation Method, and the required compatibility with the zeros of the potentials implies that the field solutions associated to this class of  models are
{\small\begin{equation}\label{solc}
\theta_{c, n}^a(x)=\tanh^{-1}\left[\cos\left( \frac{\cos^{-1}\left(\pm\tanh x\right)-(n+1/2)\pi}{a}\right)\right],  
\end{equation}}

\noindent which has the kink profile for $1\leq n< a$. So, while each $a$-value defines a new model with a finite set of topological sectors,  $n$ acts as an index for each sector. The solution (\ref{solc}) establishes which field one finds for each specified index. If we choose the negative sign for $\tanh (x)$, the asymptotic behavior of the field $\theta(x)$ obeys
\begin{equation}
\theta_{c, n}^a\left(\pm\infty\right)=\tanh^{-1}\left[\cos\left(\frac{2n\mp 1}{2}\frac{\pi}{a}\right)\right].
\end{equation}
Due to the phase $\left(n+1/2\right)\pi/a$, the kink solution acquires an asymmetric profile. The shape of the fields for all differents topological sectors in the case $a=8$ are depicted in FIG.~\ref{fig2} (we choose the case $a=8$ for all picture displayed in this work, for now on). One can see that the farther the topological sector is from the center of the potential, more and more significant is the asymmetry for the kink solution associated to the sector. In addition, the distance between two consecutive vacuums increases as they are further away from the center of the potential, which implies that the kink mass also becomes larger. So, in fact, for each $a$ one has a family of asymmetric kinks with an hierarchy on the associated masses. The same hierarchy can be observed for the topological charge $Q$. So when we choose a model with a very large value for the parameter $a$, we can obtain cases in which the kinks belong to the same model and can be found from symmetric solutions (in the center of the potential) to extremely asymmetric solutions (located in the most distant topological sectors relative to the center of potential), passing through a series of asymmetric sectors with different and increasing topological charges.

The energy density for the class of models we are studiyng in this section is given by
\begin{eqnarray}\label{edc}
\nonumber\rho_{c, n}^a(x)&=&\frac{1}{a^2}\sin ^2\left(\pi  n+\cos ^{-1}(\tanh (x))\right)\times \\
&~~&\times\csc ^2\left(\frac{\pi  n-\sin ^{-1}(\tanh (x))}{a}\right).
\end{eqnarray}
Its shape is depicted in FIG.~\ref{fig3}. Here, due to the asymmetric profile of the kinks we have, one finds also an asymmetric shape for the energy density. We can note that when $n$ approaches $a$, the asymmetry of energy density becomes more pronounced. In addition, the area under the curve decreases, which shows that the energy of the topological sector decreases as n increases.
%%%%%%%%%%%%%%%%%%%%%%%%%%%%%%%%%%%%%%%%%%%%%%

%%%%%%%%%%%%%%%%%%%%%%%%%%%%%%%%%%%%%%%%%%%%%%
%%%%%%%%%%%%%%%%%%%%%%%%%%%%%%%%%%%%%%%%%%%%%%
\begin{figure}[t]
\renewcommand{\thefigure}{4}
\centerline{\includegraphics[height=15em]{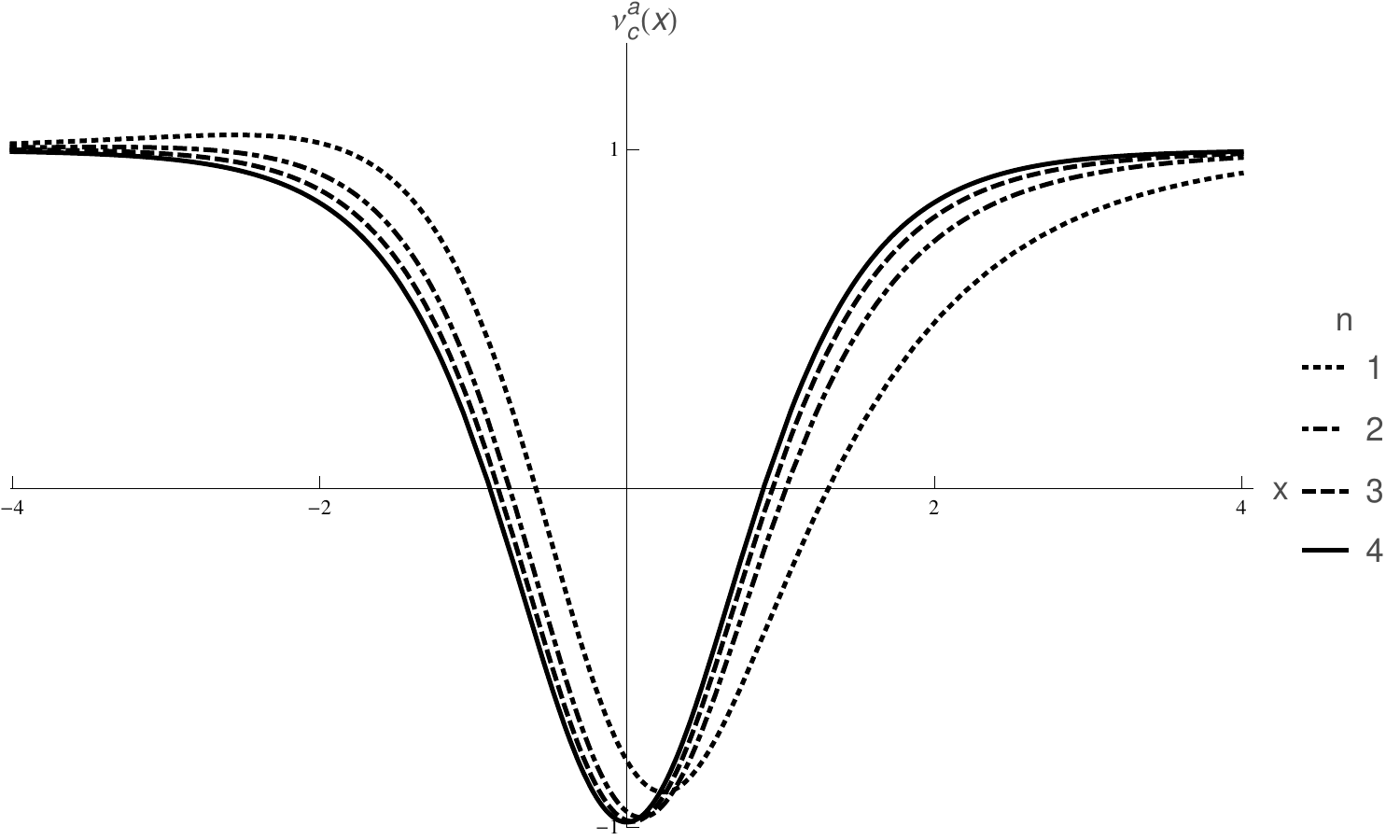}}
\caption{The stability potential (\ref{spc}) for $a=8$ and $n=1,2,3$ and 4.}\label{fig4}
\end{figure}
%%%%%%%%%%%%%%%%%%%%%%%%%%%%%%%%%%%%%%%%%%%%%%
%%%%%%%%%%%%%%%%%%%%%%%%%%%%%%%%%%%%%%%%%%%%%%
%%%%%%%%%%%%%%%%%%%%%%%%%%%%%%%%%%%%%%%%%%%%%%
%%%%%%%%%%%%%%%%%%%%%%%%%%%%%%%%%%%%%%%%%%%%%%
\begin{figure}[t]
\renewcommand{\thefigure}{5}
\centerline{\includegraphics[height=15em]{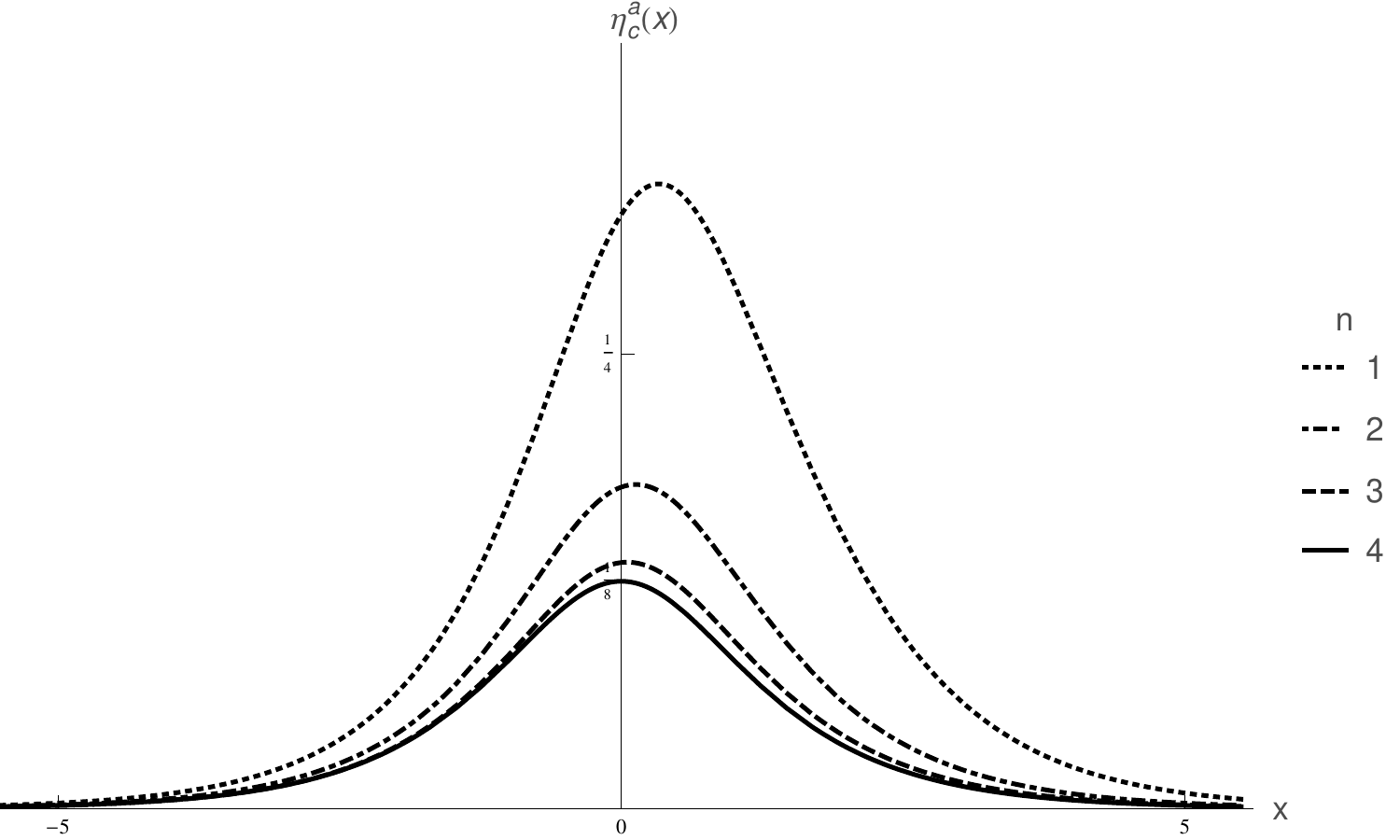}}
\caption{The zero mode (\ref{zmc}) for $a=8$ and $n=1,2,3$ and 4.}\label{fig5}
\end{figure}
%%%%%%%%%%%%%%%%%%%%%%%%%%%%%%%%%%%%%%%%%%%%%%
%%%%%%%%%%%%%%%%%%%%%%%%%%%%%%%%%%%%%%%%%%%%%%

For the stability potential, one has the expression (\ref{stabilitypotential}), here represented by
%%%%%%%%%%%%%%%%%%%%%%%%%%%%%%%%%%%%%%%%%
%%%%%%%%%%%%%%%%%%%%%%%%%%%%%%%%%%%%%%%%%
\begin{eqnarray}\label{spc}
\nonumber v_{c, n}^a(x)&=&\frac{1}{a^2}\left[\mathcal{T}_a\left(\mathbb{F}_{c, n}(x)\right){}^2 \left(\cosh \left(2  \tanh^{-1}\left(\mathbb{F}_{c, n}(x)\right)\right)-\right.\right. \\%
\nonumber &~&-\left. a^2\right)+ a^2 \left(1-\mathbb{F}_{c, n}(x)^2\right)\mathcal{U}_{a-1}\left(\mathbb{F}_{c, n}(x)\right){}^2+\\
&~& +3a \mathbb{F}_{c, n}(x) \mathcal{T}_a\left(\mathbb{F}_{c, n}(x)\right)\left. \mathcal{U}_{a-1}\left(\mathbb{F}_{c, n}(x)\right)\right],
\end{eqnarray}
%%%%%%%%%%%%%%%%%%%%%%%%%%%%%%%%%%%%%%%%%%%
%%%%%%%%%%%%%%%%%%%%%%%%%%%%%%%%%%%%%%%%%%%
with $\mathbb{F}_{c, n}(x)=\cos \left(\frac{\sin ^{-1}(\tanh (x))-\pi  n}{a}\right)$. It is displayed in FIG.~\ref{fig4}. In this case the asymmetric profile can be observed, leading to a deformed reflectionless modified Poschl-Teller-like stability potential. As we keep the translational invariance of the solutions, the zero-mode exists and is given by
\begin{equation}\label{zmc}
\eta_{c, n}^a(x)=\frac{1}{a}\text{sech}(x) \csc \left(\frac{\pi  n-\sin ^{-1}(\tanh (x))}{a}\right),
\end{equation}
as shown in FIG.~\ref{fig5}.
\subsection{Model 2}
%%%%%%%%%%%%%%%%%%%%%%%%%%%%%%%%%%%%%%%%%%%%%%
%%%%%%%%%%%%%%%%%%%%%%%%%%%%%%%%%%%%%%%%%%%%%%
In this section we are interested in studying the second type of models present in this work, which are defined for $m=0,\pm 1,\pm 2, ...,$ i.e., 
\begin{equation}\label{modelu}
u_s^a\left(\theta\right)=\frac{1}{2a^2}\mathcal{U}_{a-1}^2\left(\tanh\theta\right).
\end{equation}
In this case, we have potentials with $\mathcal{Z}_2$-symmetry and nontrivial internal structure. The models represented by (\ref{modelu}) are  bounded in the interval $[0,1/2)$, since for $\theta\rightarrow\infty$ we have $\mathcal{U}_{a-1}(\tanh\theta)\rightarrow\mathcal{U}_{a-1}(1)=a$, implying that $ u_s^a(\pm\infty)=1/2 $.   It is an interesting fact, because the models described by (\ref{modelu}) describe physical situations where the systems present asymptotic behavior that are different from the previous models. Thus, it is clear that despite the fact that (\ref{modelu}) and (\ref{potc}) are derived from the same deformation process, they have relevant qualitative differences. 

\begin{figure}[t]
\renewcommand{\thefigure}{6.1}
\centerline{\includegraphics[height=15em]{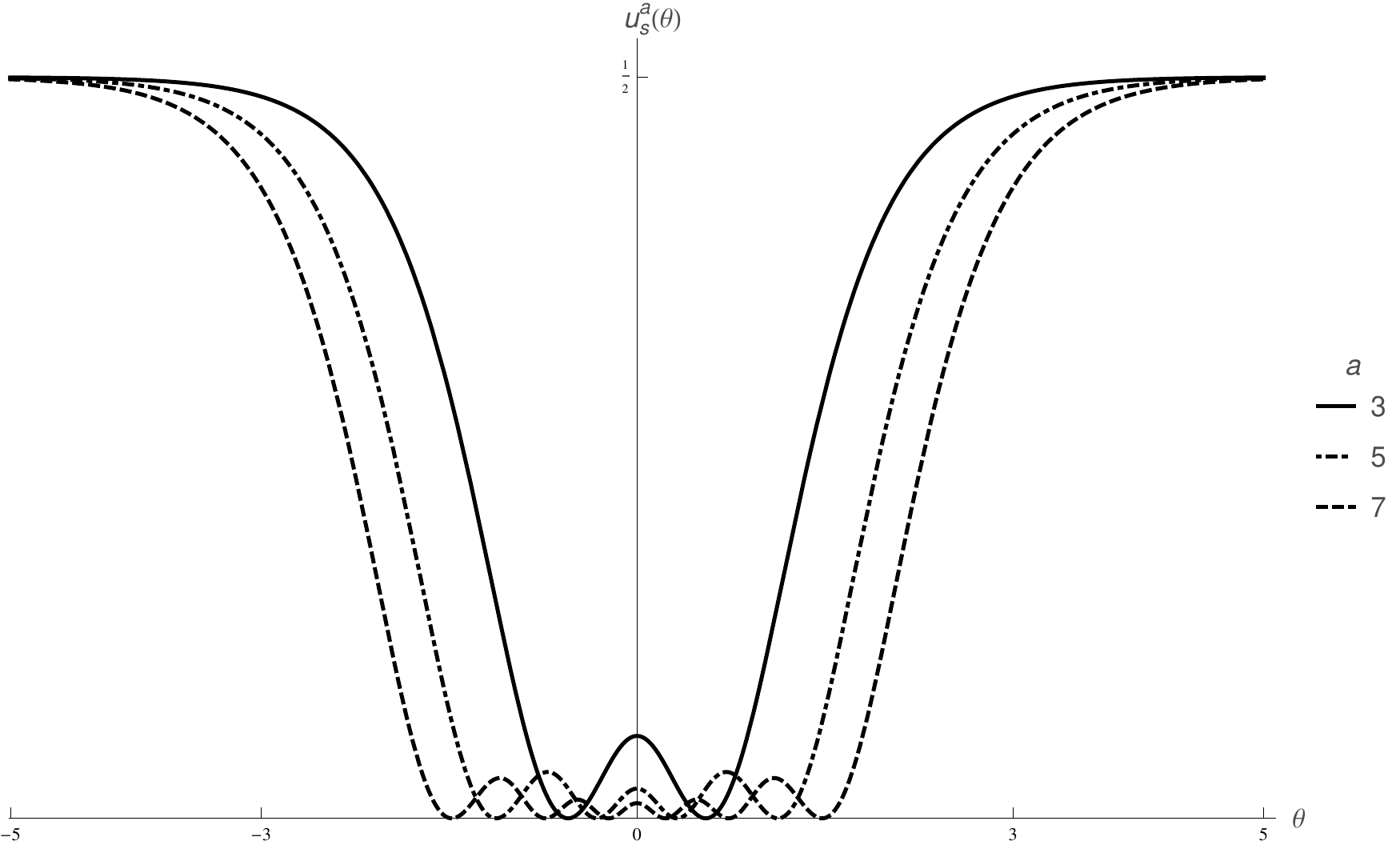}}
\caption{Potential $u_s^a\left(\theta\right)$ for some values of odd $a$.}\label{fig6.1}
\end{figure}

\begin{figure}[t]
\renewcommand{\thefigure}{6.2}
\centerline{\includegraphics[height=15em]{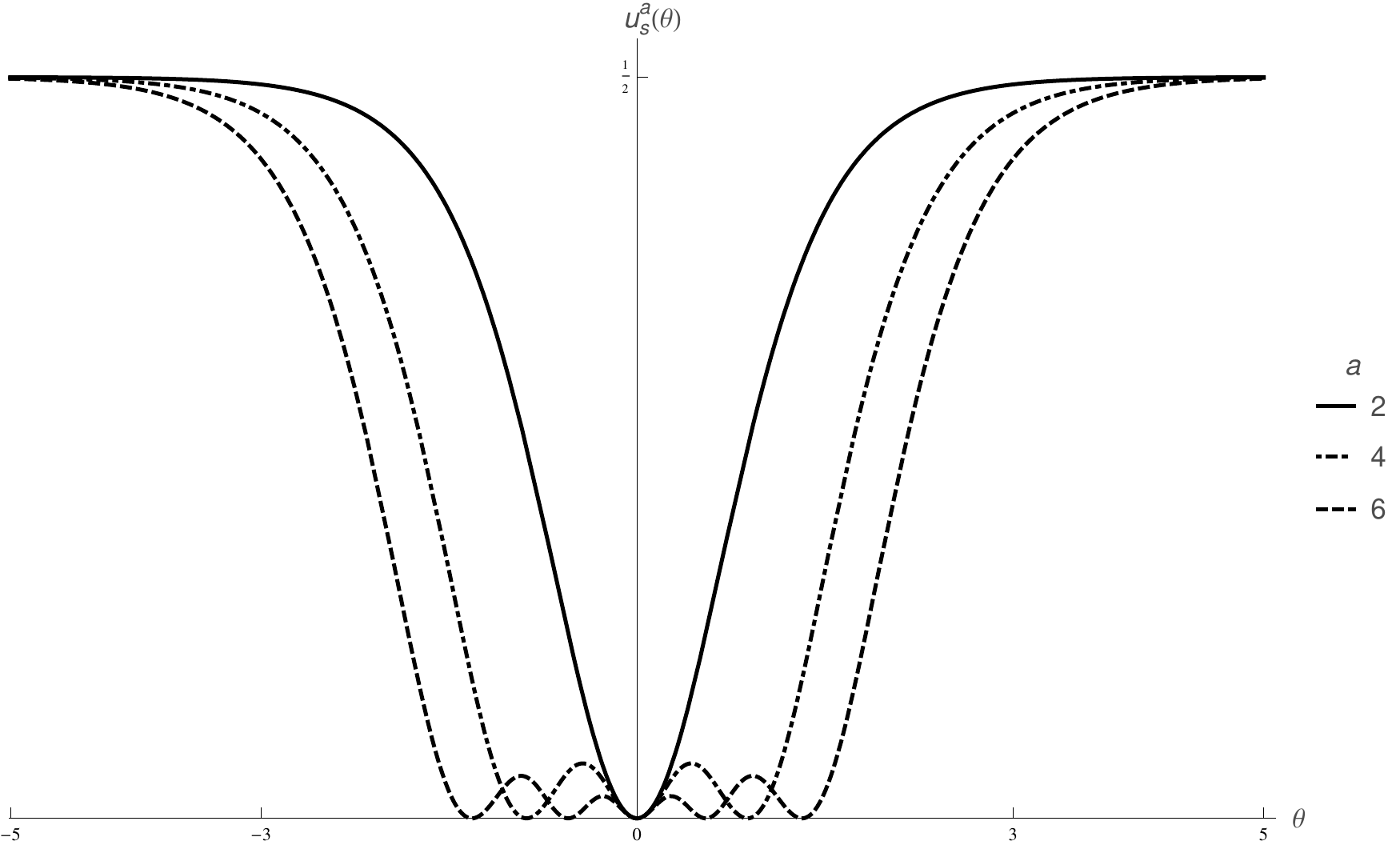}}
\caption{Potential $u_s^a\left(\theta\right)$ for some values of even $a$.}\label{fig6.2}
\end{figure}

\begin{figure}[t]
\renewcommand{\thefigure}{7}
\centerline{\includegraphics[height=15em]{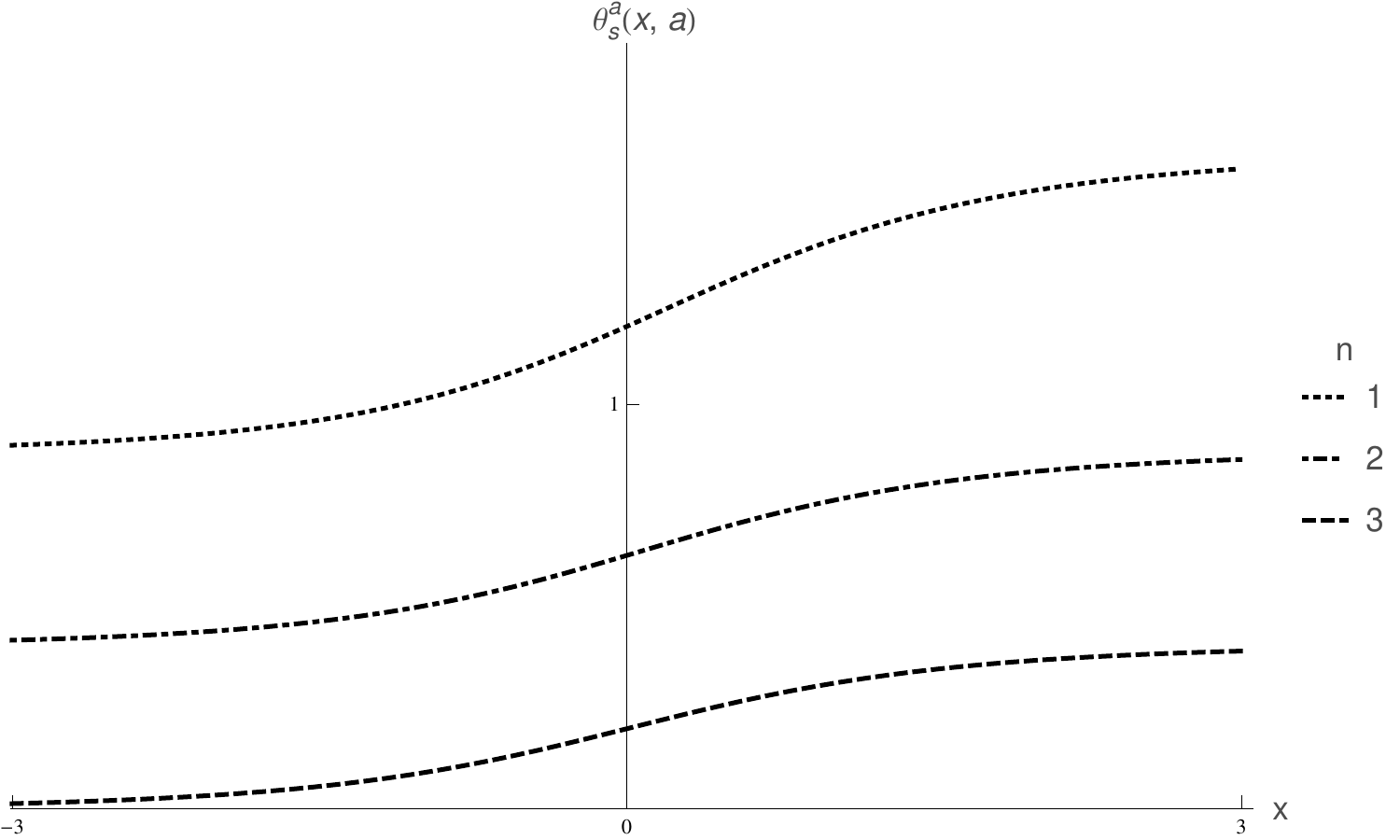}}
\caption{The field solutionl (\ref{sols}) for $a=8$ and $n=1,2,$ and 3}\label{fig7}
\end{figure}

The first potentials of the class represented by (\ref{modelu}) are given by 
\begin{eqnarray}
\nonumber u_s^2\left(\theta\right)&=&\frac{1}{2}\tanh ^2\theta \\
\nonumber u_s^3\left(\theta\right)&=&\frac{1}{18} \left(4 \tanh ^2\theta -1\right)^2\\
\nonumber u_s^4\left(\theta\right)&=&\frac{1}{32} \left(8 \tanh ^3\theta -4 \tanh \theta \right)^2\\
u_s^5\left(\theta\right)&=&\frac{1}{50} \left(16 \tanh ^4\theta-12 \tanh ^2\theta+1\right)^2
%\nonumber u_s^5\left(\theta\right)&=&\frac{1}{50} \left(16 \tanh ^4\theta -12 \tanh ^2\theta +1\right)^2\\
%\nonumber u_s^6\left(\theta\right)&=&\frac{1}{72} \left(32 \tanh ^5\theta -32 \tanh ^3\theta +6 \tanh \theta \right)^2\\
%\nonumber u_s^7\left(\theta\right)&=&\frac{1}{98} \left(64 \tanh ^6\theta -80 \tanh ^4\theta +24 \tanh ^2\theta-1\right)^2
\end{eqnarray}
Now  we have $(a-1)/2$ different topological sectors for odd $a$ and $(a-2)/2$ different topological sectors for even $a$,  and its zeros are distributed  in discrete values by formula $\theta_{s,n}=\tanh^{-1}\left[\cos\left(\frac{n\pi}{a}\right)\right]$, for $0< n< a.$ These models are depicted in FIG.~\ref{fig6.1} for odd $a$ and in FIG.~\ref{fig6.2} for even $a$. From Trigonometry, for each $n$ we have $\cos \left(n\pi/a \right)=\cos\left((a-n)\pi/a \right)$, which reflects de $Z_2$ symmetry present in the model. As it happens with Model 1, we can observe that as long as $a$ increases, the number of topological sectors also gets bigger.  We can also find the $W$-function for each model we use. For example, the first models are derived from the functions
\begin{eqnarray}
W_s^2(\theta)&=&\log (\cosh \theta)+c_2\\
\nonumber W_s^3(\theta)&=&\theta-\frac{4}{3} \tanh \theta+c_3\\
\nonumber W_s^4(\theta)&=&\log (\cosh \theta)+\text{sech}^2\theta+c_4\\
\nonumber W_s^5(\theta)&=&\theta+\frac{4}{15} \tanh \theta \left(4 \text{sech}^2 \theta-7\right)+c_5
\end{eqnarray}
%%%%%%%%%%%%%%%%%%%%%%%%%%%%%%%%%%%%%%%%%%%%%%
%%%%%%%%%%%%%%%%%%%%%%%%%%%%%%%%%%%%%%%%%%%%%%
%%%%%%%%%%%%%%%%%%%%%%%%%%%%%%%%%%%%%%%%%%%%%%
%%%%%%%%%%%%%%%%%%%%%%%%%%%%%%%%%%%%%%%%%%%%%%
%%%%%%%%%%%%%%%%%%%%%%%%%%%%%%%%%%%%%%%%%%%%%%%%
%%%%%%%%%%%%%%%%%%%%%%%%%%%%%%%%%%%%%%%%%%%%%%%%
%%%%%%%%%%%%%%%%%%%%%%%%%%%%%%%%%%%%%%%%%%%%%%
%%%%%%%%%%%%%%%%%%%%%%%%%%%%%%%%%%%%%%%%%%%%%%
%%%%%%%%%%%%%%%%%%%%%%%%%%%%%%%%%%%%%%%%%%%%%%
%%%%%%%%%%%%%%%%%%%%%%%%%%%%%%%%%%%%%%%%%%%%%%
 Note that in the center of the potencial one has $$u_s^a(0)=\left\{
\begin{array}{cc}
0&\text{, if $a$ is even}\\
\frac{1}{2a^2}&\text{, if $a$ is odd} .\\
\end{array}
\right.$$
Thus, for odd values of $ a $ we have a topological sector passing through the potential center, whereas this does not happen for even values of $ a $. %%%%%%%%%%%%%%%%%%%%%%%%%%%%%%%%%%%%%%%%%%%%%%%%
%%%%%%%%%%%%%%%%%%%%%%%%%%%%%%%%%%%%%%%%%%%%%%%%
The differential equation that we have to solve now is
\begin{equation}\label{1eq}
\theta'= \frac{1}{a}\mathcal{U}_{a-1}\left(\tanh\theta\right).
\end{equation}
%%%%%%%%%%%%%%%%%%%%%%%%%%%%%%%%%%%%%%%%%%%%%%%%%
%%%%%%%%%%%%%%%%%%%%%%%%%%%%%%%%%%%%%%%%%%%%%%%%%
The solution is again given by (\ref{gensolu}), but now the compatibility between the minima of the potential and the asymptotic boundaries of the fields imply a different solution. In particular we must have $m=-(n+1)$. So, the complete solution is
\begin{equation}\label{sols}
\theta_{s, n}^a(x)=\tanh^{-1} \left[\cos\left( \frac{\cos^{-1}\left(-\tanh x\right)-(n+1)\pi}{a}\right)\right],
\end{equation}
The above field has kinks when $ 0 < n <a-1 $. There is also a degeneracy for $ n $ and $ n \pm 2ka $. The case $ a = 8 $ is shown in figure FIG. \ref {fig7}. Note that, unlike the first model, we now have only three topological solutions for this specific case of field solutions. In this case the asymptotics  of the field for each kink are
\begin{eqnarray}
\nonumber \theta_{s, n}^a(\infty)&=&\tanh^{-1}\left[\cos\left( \frac{n\pi}{a}\right)\right]\\
\theta_{s,  n}^a(-\infty)&=&\tanh^{-1}\left[\cos\left( \frac{n+1}{a}\pi\right)\right].
\end{eqnarray}
Thus, as in the first model, we have a class of asymmetric kinks in which the asymmetry becomes increasingly more or less evident as far as the topological sector is located farther or near the center of potential. We also observe here a hierarchy of charges due to the fact that the further a sector is from the center of potential, the greater the distance between two consecutive minima.%%%%%%%%%%%%%%%%%%%%%%%%%%%%%%%%%%%%%%%%%%%%%%
%%%%%%%%%%%%%%%%%%%%%%%%%%%%%%%%%%%%%%%%%%%%%%
%%%%%%%%%%%%%%%%%%%%%%%%%%%%%%%%%%%%%%%%%%%%%%
%%%%%%%%%%%%%%%%%%%%%%%%%%%%%%%%%%%%%%%%%%%%%%

\begin{figure}[t]
\renewcommand{\thefigure}{8}
\centerline{\includegraphics[height=15em]{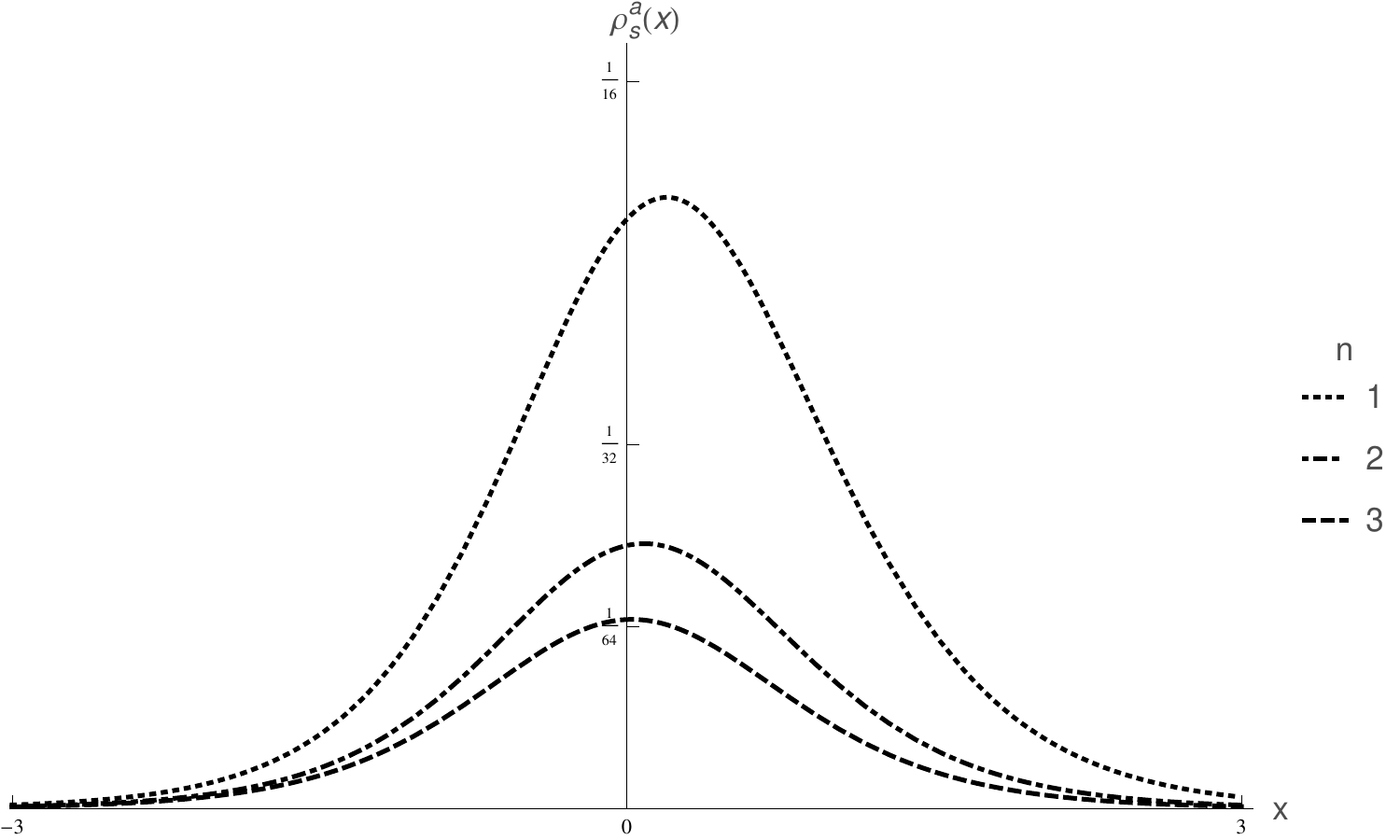}}
\caption{The energy density (\ref{eds}) for $a=8$ and $n=1,2,$ and 3}\label{fig8}
\end{figure}

The energy density of the model we study in this section is given by the function
\begin{equation}\label{eds}
\rho_{s, n}^a(x)=\frac{1}{a^2}\mathcal{U}_{a-1}^2\left(\cos\left(\frac{\cos^{-1}\left(\tanh x\right)-n\pi }{a}\right)\right),
\end{equation}
and its form for the case $a=8 $ is represented in FIG.~\ref{fig8}. Note that the asymmetric profile present in energy density (\ref{eds}) has qualitative similarities with energy density (\ref{edc}), but now the growth of asymmetry in relation to $ n $ is less accentuated. Another relevant difference occurs in the behavior of the field solutions for $n = -1, 0, a-1$ and $a$. They have divergent energy, but now instead of exploding into infinity, they have in one of their limits $\phi (x\rightarrow \pm\infty)\rightarrow constant$, while in the other it has a linear divergence, $\phi (x \sim\mp\infty)\sim \alpha x +\beta$. This is due to the {\it plat\^o} present in the potential of Model 2. It is clear in the equation (\ref{1order}) that if the potential approaches asymptotically to a constant value, then the derivative of the field also approaches a constant. Thus an asymptotic field should approach a straight line.
%%%%%%%%%%%%%%%%%%%%%%%%%%%%%%%%%%%%%%
%%%%%%%%%%%%%%%%%%%%%%%%%%%%%%%%%%%%%%
%%%%%%%%%%%%%%%%%%%%%%%%%%%%%%%%%%%%%%%%%%%%%%
%%%%%%%%%%%%%%%%%%%%%%%%%%%%%%%%%%%%%%%%%%%%%%

%%%%%%%%%%%%%%%%%%%%%%%%%%%%%%%%%%%%%%%%%%%%%%%%%
%%%%%%%%%%%%%%%%%%%%%%%%%%%%%%%%%%%%%%%%%%%%%%%%%

\begin{figure}[t]
\renewcommand{\thefigure}{9}
\centerline{\includegraphics[height=15em]{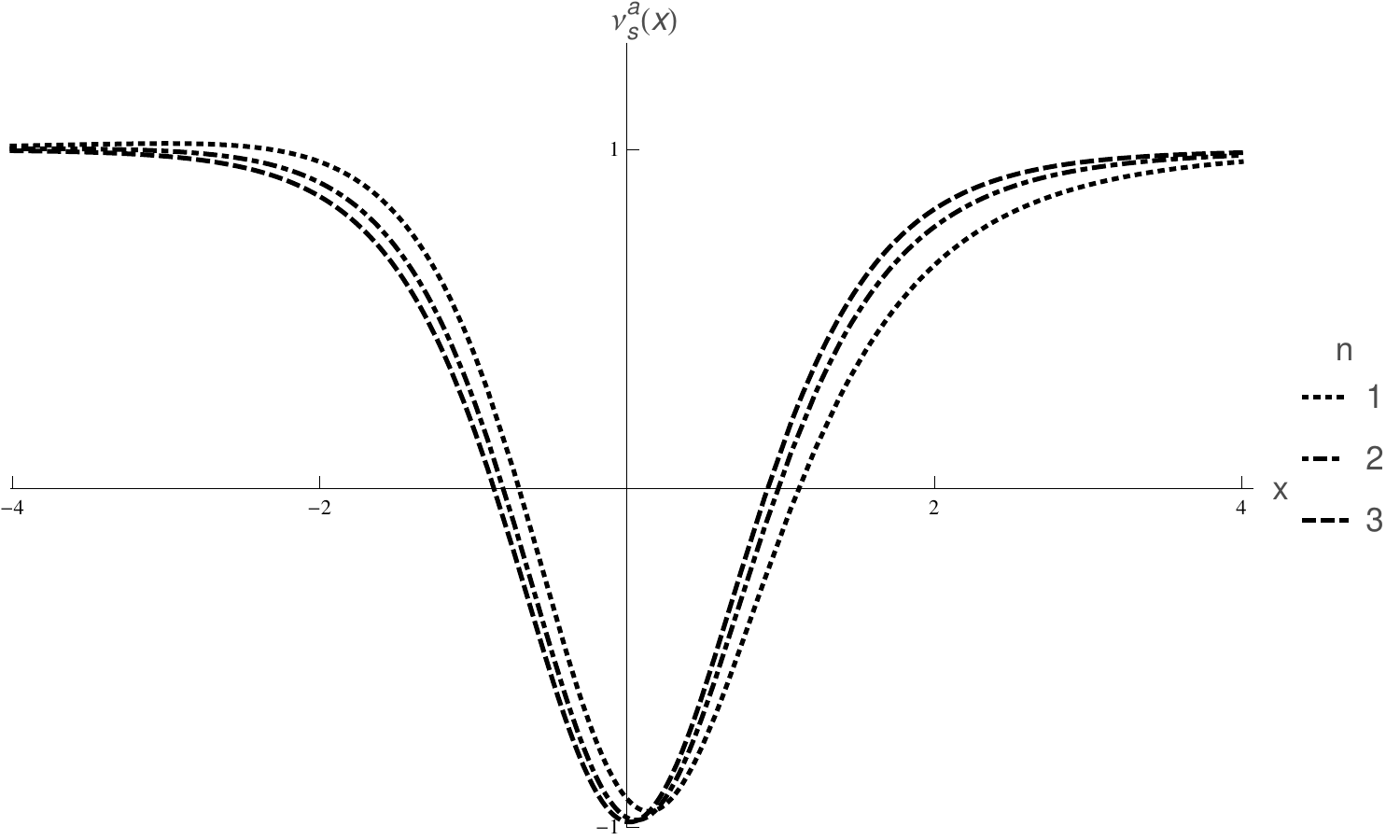}}
\caption{The stability potential (\ref{sps}) for $a=8$ and $n=1,2,$ and 3}\label{fig9}
\end{figure}

\begin{figure}[t]
\renewcommand{\thefigure}{10}
\centerline{\includegraphics[height=15em]{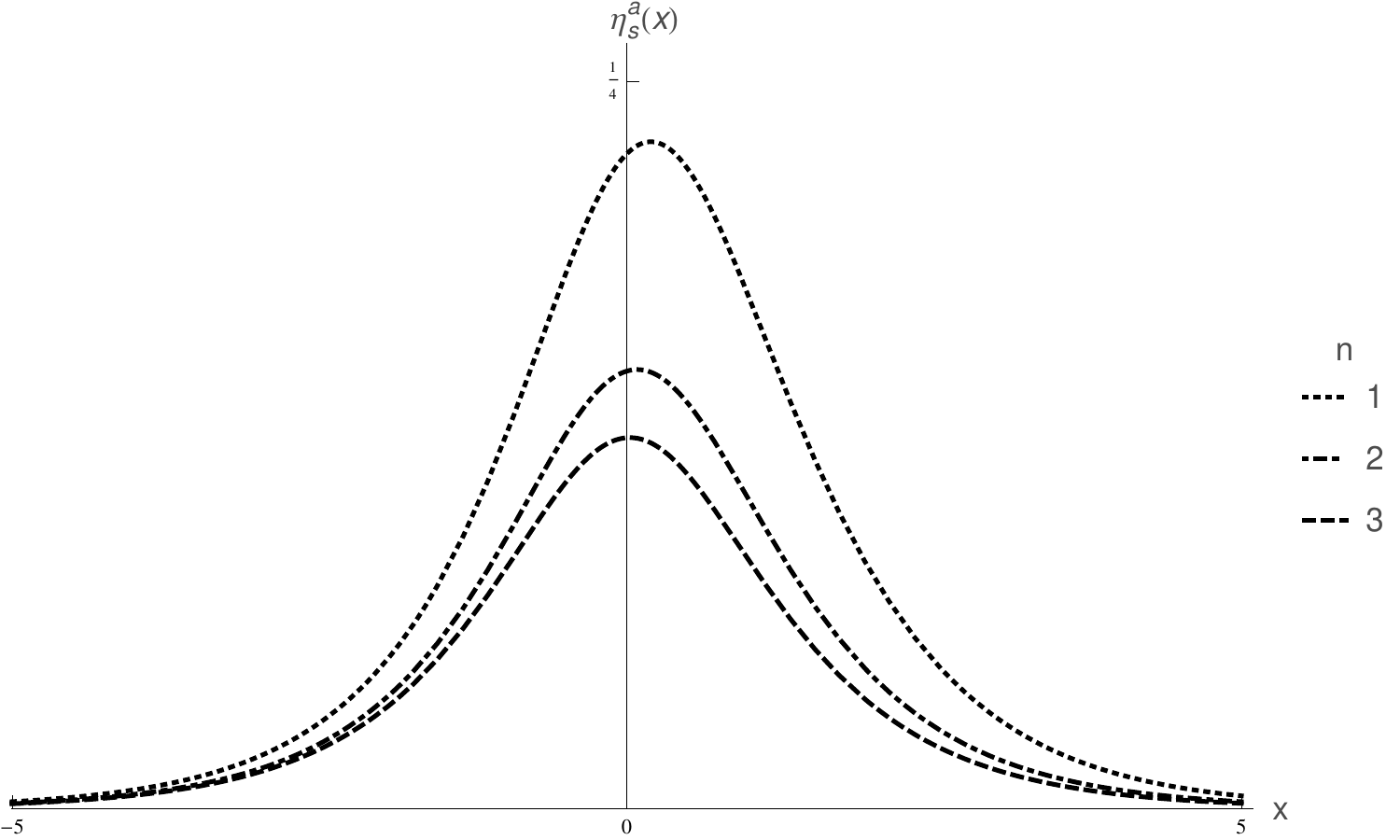}}
\caption{The zero mode (\ref{zms}) for $a=8$ and $n=1,2,$ and 3}\label{fig10}
\end{figure}

The stability potential  derived from (\ref{modelu}) is given by the expression
\begin{eqnarray}\label{sps}
\nonumber v_{s, n}^a(x)&=&\left[\mathcal{T}_a\left(\mathbb{F}_{s,n}(x)\right)-\frac{\mathbb{F}_{s,n}(x) \mathcal{U}_{a-1}\left(\mathbb{F}_{s,n}(x)\right)}{a}\right]^2\\
\nonumber &{}&+\frac{1}{a}\mathcal{U}_{a-1}\left(\mathbb{F}_{s,n}(x)\right) \left[\mathbb{F}_{s,n}(x) U_{a-2}\left(\mathbb{F}_{s,n}(x)\right)-\right. \\
\nonumber &{}&\left. -\frac{(a-1)}{2a} \left(1-\mathbb{F}_{s,n}(x)^2\right) \mathcal{U}_{a-1}\left(\mathbb{F}_{s,n}(x)\right)\right.\times \\
&{}&\times \left.\left(\cosh \left(2 \tanh ^{-1}\left(\mathbb{F}_{s,n}(x)\right)\right)+2 a+1\right)\right].\;\;\;
\end{eqnarray}
Here $\mathbb{F}_{s,n}(x)=\cos \left(\frac{\cos ^{-1}(\tanh (x))-\pi  n}{a}\right)$. It is shown in FIG.~\ref{fig9}. The zero mode has the form
\begin{equation}\label{zms}
\eta^a_{s,n}= -\frac{1}{a}\text{sech}(x) \csc \left(\frac{\pi  n-\cos ^{-1}(\tanh (x))}{a}\right)
\end{equation}
and it is show in FIG.~\ref{fig10}. These quantities show more clearly the qualitative similarities between the models, although we have seen that they describe different systems.
%%%%%%%%%%%%%%%%%%%%%%%%%%%%%%%%%%%%%%%%%%%%%%
%%%%%%%%%%%%%%%%%%%%%%%%%%%%%%%%%%%%%%%%%%%%%%
%%%%%%%%%%%%%%%%%%%%%%%%%%%%%%%%%%%%%%%%%%%%%%
%%%%%%%%%%%%%%%%%%%%%%%%%%%%%%%%%%%%%%%%%%%%%%
%%%%%%%%%%%%%%%%%%%%%%%%%%%%%%%%%%%%%%%%%%%%%%
%%%%%%%%%%%%%%%%%%%%%%%%%%%%%%%%%%%%%%%%%%%%%%
\section{Thick Branes}
%%%%%%%%%%%%%%%%%%%%%%%%%%%%%%%%%%%%%%%%%%%%%%
%%%%%%%%%%%%%%%%%%%%%%%%%%%%%%%%%%%%%%%%%%%%%%
Field Theory models also motivate the study of new scenarios with thick branes \cite{gw, fre, csaki, gremm, bm}. In this perspective, the scalar field acts as a source of gravity and models the way gravity is distributed around the brane. Here we are interested in the system where we have a 3-brane embedded in a (4 + 1) space-time with an extra dimension of infinite extent. The background geometry is represented by a static metric given by
\begin{equation}\label{metric}
ds^2_5=g_{ab}dx^a dx^b=e^{2A(y)}ds^2_4-dy^2.
\end{equation}
Where $a,  b=0,..., 4$, $\mu, \nu=0,..., 3$, $ds^2_4=\eta_{\mu\nu}dx^{\mu}dx^{\nu}$, the $y$-coordinate describes the extra spatial dimension, $A(y)$ is the warp function, which we assume to be dependent only of the extra dimension and $e^{A(y)}$ is the warp factor. 

\begin{figure}[t]
\renewcommand{\thefigure}{11}
\centerline{\includegraphics[height=15em]{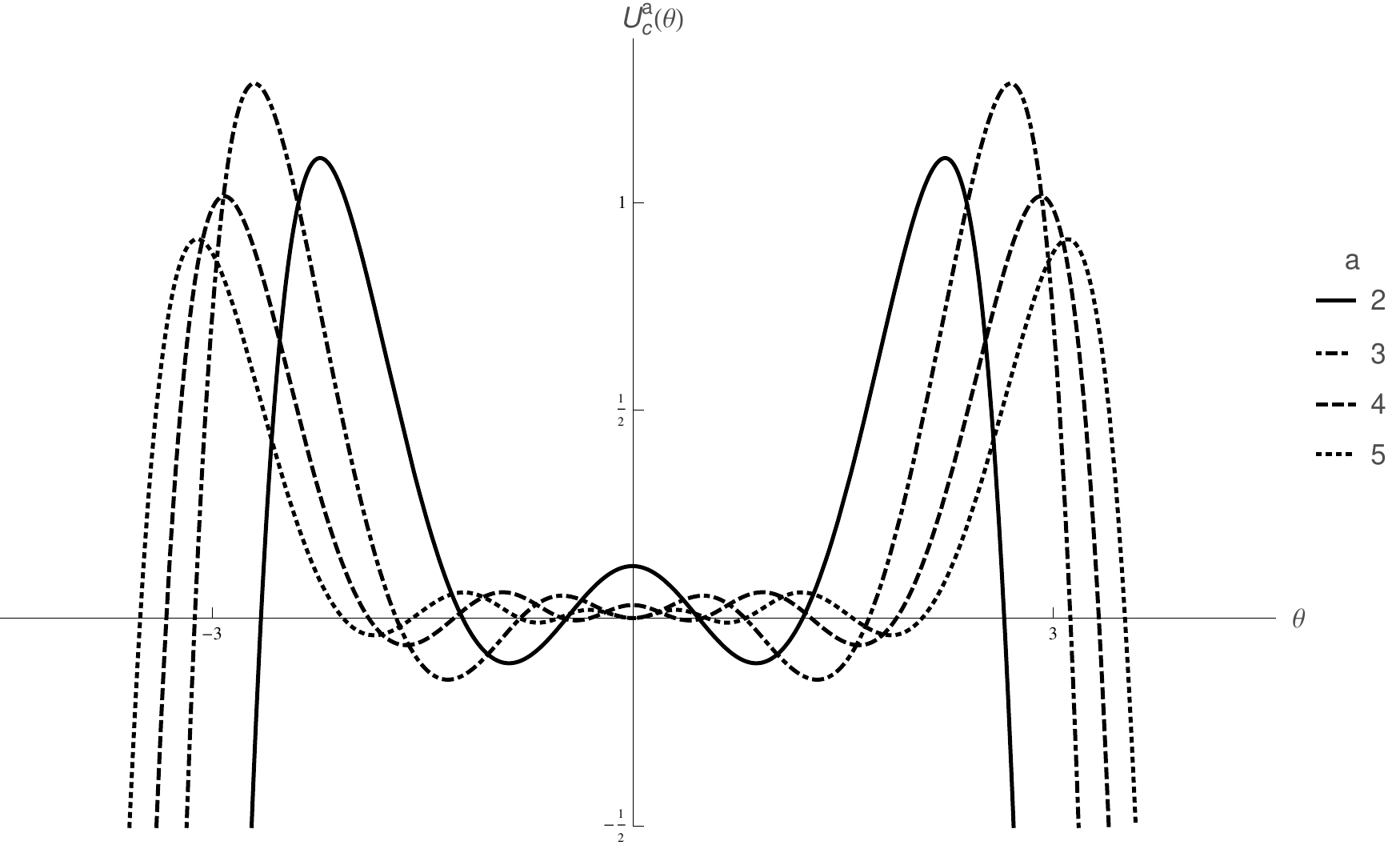}}
\caption{Potential (\ref{poten}) for Model 1 with $a=2, 3, 5$ and $8$. In this picture, we made a fine-tunning in the constants $c_i's$ such that $W_{c,a}(0)=0$ }\label{fig11}
\end{figure}

The braneworld model that we work here is described by the action
\begin{equation}\label{action}
S=\int d^5x\sqrt{|g|}\left(-\frac{1}{4}R+\mathcal{L}\right).
\end{equation}
Here, $\mathcal{L}(\phi,\partial_{a}\phi)=\frac{1}{2}g_{ab}\partial^{a}\phi\partial^{b}\phi-U(\phi)$  and, for simplicity, $4\pi G_5=1$. We also assume that the field also depends only of the extra dimension, i.e., $\phi=\phi(y)$. The action (\ref{action}) leads us to the Einstein equations
\begin{equation}
G_{ab}=2T_{ab}.
\end{equation}
The equations for $00$ and $44$ components are, respectively, 
\bes\label{29}\begin{eqnarray}
 6A'^2&=&\phi'^2-2U,\\
3A''+6A'^2&=&-\phi^{\prime 2}-2 U.
\end{eqnarray}\ees
Here, the prime representes differentiation with respect to the coordinate $y$. We can rewrite \eqref{29}, subtracting the first equation from the second one, to obtain
\begin{equation}\label{asecondorder}
A''=-\frac{2}{3}\phi'^2. 
\end{equation}
The equation above allows us to rewrite the model in terms of first order equations. The W-function is inserted in the system by the choice
\begin{equation}\label{afirstorder}
A'=-\frac{2}{3}\,W(\phi(y)),
\end{equation}
and it implies that the field now is obtained from the equation  $\phi'=W_\phi$.  In this section, we want to study the thick brane scenarios generated by the models analyzed in the previous sections. We'll just makes the analysis for model 1, because qualitatively the model 2 is very similar. To solve Einstein equations (\ref{29}), the potential for the model under investigation must be given by the expression
\be\label{poten}
U(\phi)=\frac12\,W_\phi^2-\frac43\, W^2.
\ee
We do not have an analytical form for the W function of the models we are working on here, but we can plot their figure for some cases, which are shown in Fig. \ref{fig11} and one can see that these models have symmetry $\mathcal{Z } _2$. The solution of the field, as can be seen, is given by the equation (\ref{solc}), through the permutations $\theta \rightarrow \phi $ e $ x \rightarrow y $. So, in fact, for every $ a $ we have a set of different thick and asymmetric branes.
%%%%%%%%%%%%%%%%%%%%%%%%%%%%%%%%%%%%%%%%%%%%%%
%%%%%%%%%%%%%%%%%%%%%%%%%%%%%%%%%%%%%%%%%%%%%%
%%%%%%%%%%%%%%%%%%%%%%%%%%%%%%%%%%%%%%%%%%%%%%
%%%%%%%%%%%%%%%%%%%%%%%%%%%%%%%%%%%%%%%%%%%%%%
\begin{figure}[t]
\renewcommand{\thefigure}{12}
\centerline{\includegraphics[height=15em]{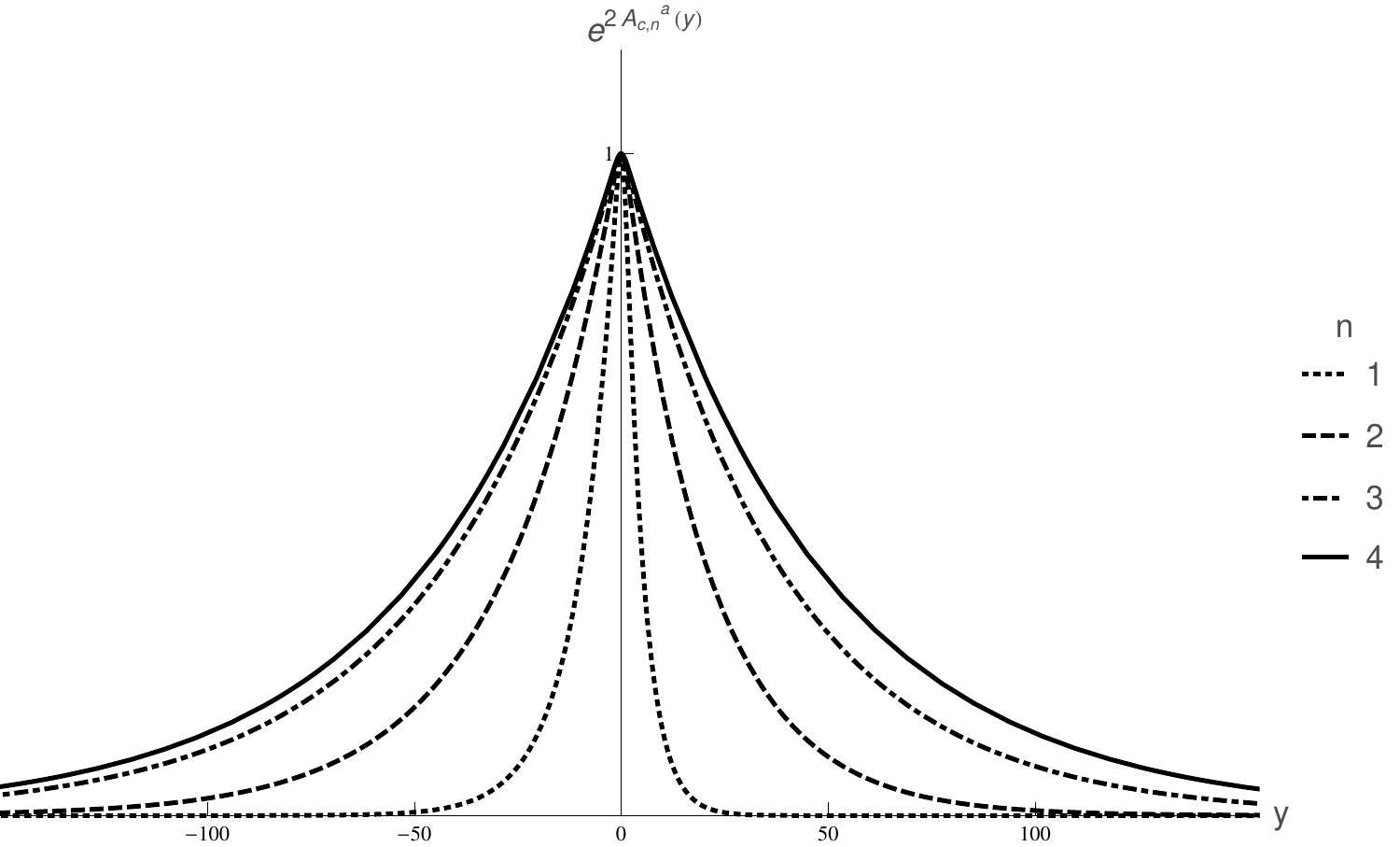}}
\caption{The warp function for $a=8$ and $n=1,2,$ and 3}\label{fig12}
\end{figure}
%%%%%%%%%%%%%%%%%%%%%%%%%%%%%%%%%%%%%%%%%%%%%%
%%%%%%%%%%%%%%%%%%%%%%%%%%%%%%%%%%%%%%%%%%%%%%

The energy density derived from the energy-momentum tensor is
\be\label{edb}
\rho(y)= e^{2 A}\left(\frac12\phi'^2+U(\phi)\right)
\ee 
and models derived from the equation (\ref{afirstorder}) have zero total energy. This is because the energy density can be rewritten as $\rho(y)=\frac{d}{dy}\left(W\,e^{2A}\right)$. An implicit requirement for the W function is that it must be finite in the coordinate space when $ y \rightarrow \pm\infty$. This is necessary so that the BPS-energy (\ref{Ebps}) of the system is finite. On the other hand, the warp function must be an decreasing function in the bulk. The shape of $e^{2A(y)}$ is depicted in FIG.~(\ref{fig12}), where we solve (\ref{asecondorder}) with boundary conditions $A(0)=A'(0)=0$. 

\begin{figure}[ht]
\renewcommand{\thefigure}{13}
\centerline{\includegraphics[height=16em]{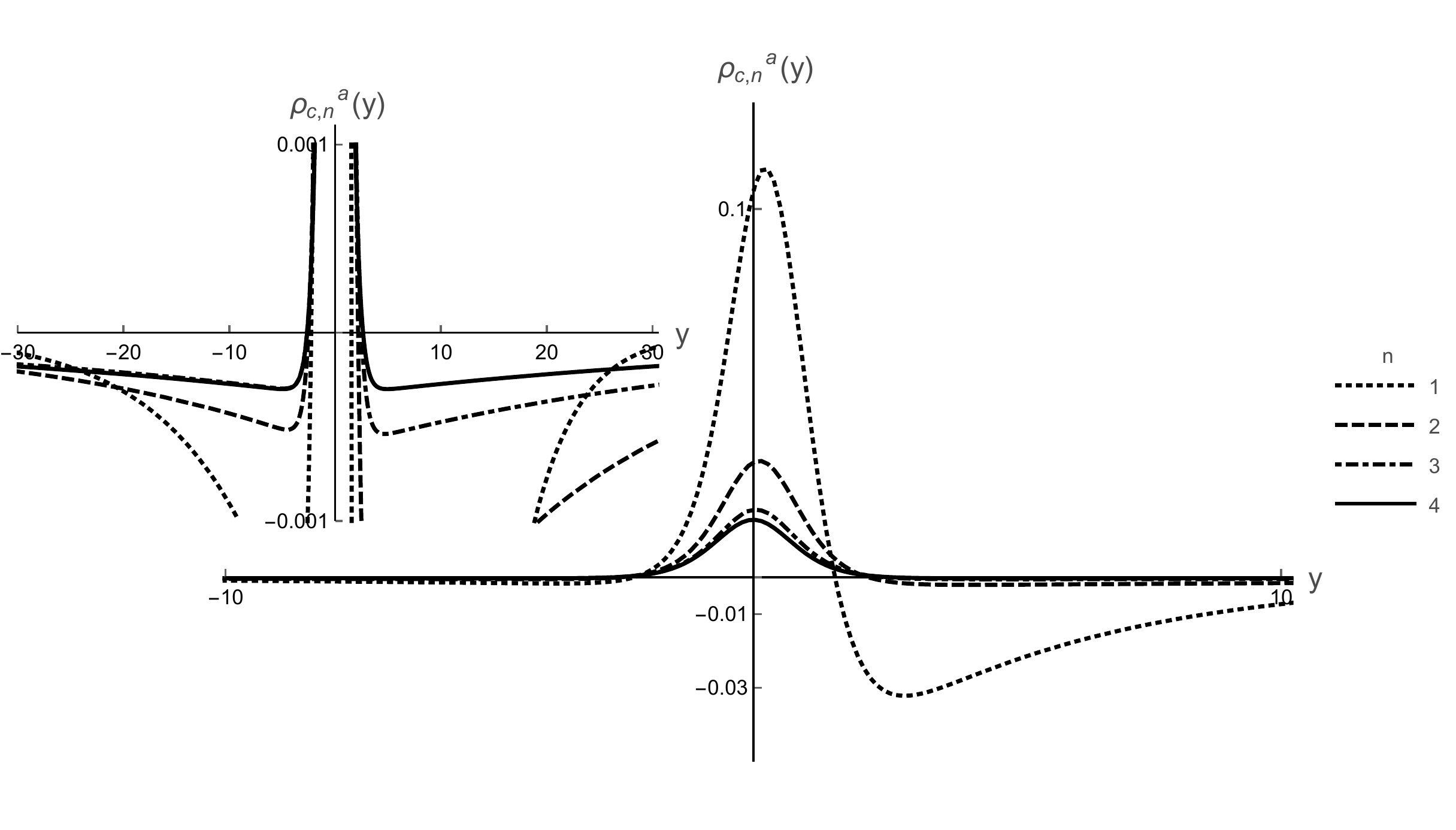}}
\caption{The energy density (\ref{edb}) for $a=8$ and $n=1,2,3$ and 4. The bigger picture is plotted in the range (-10,10), and the smaller is a zoom in the range (-30,30).}\label{fig13}
\end{figure}

In this case we have $1 \leq n <a$, except for a change of variables  $y \rightarrow -y$. We can have $ a/2 $ branes if $a$ is even or $ (a-1)/2 $ branes if $a$ is odd. When the values ​​of $n$ decrease, we can observe that the brane of the model becomes thinner and with a more visible asymmetry, so we expect the energy density to be more concentrated on one side of the brane, as shown in FIG.(\ref{fig13}). It reflects the fact that the cost of energy is higher when gravity is confined to regions closer to the source.
%\onecolumngrid
%begin{widetext}
%%%%%%%%%%%%%%%%%%%%%%%%%%%%%%%%%%%%%%%%%%%%%%
%%%%%%%%%%%%%%%%%%%%%%%%%%%%%%%%%%%%%%%%%%%%%%
%%%%%%%%%%%%%%%%%%%%%%%%%%%%%%%%%%%%%%%%%%%%%%
%%%%%%%%%%%%%%%%%%%%%%%%%%%%%%%%%%%%%%%%%%%%%%
%\end{widetext}

%%%%%%%%%%%%%%%%%%%%%%%%%%%%%%%%%%%%%%%%%%%%%%
\begin{figure}[t]
\renewcommand{\thefigure}{14}
\centerline{\includegraphics[height=12em]{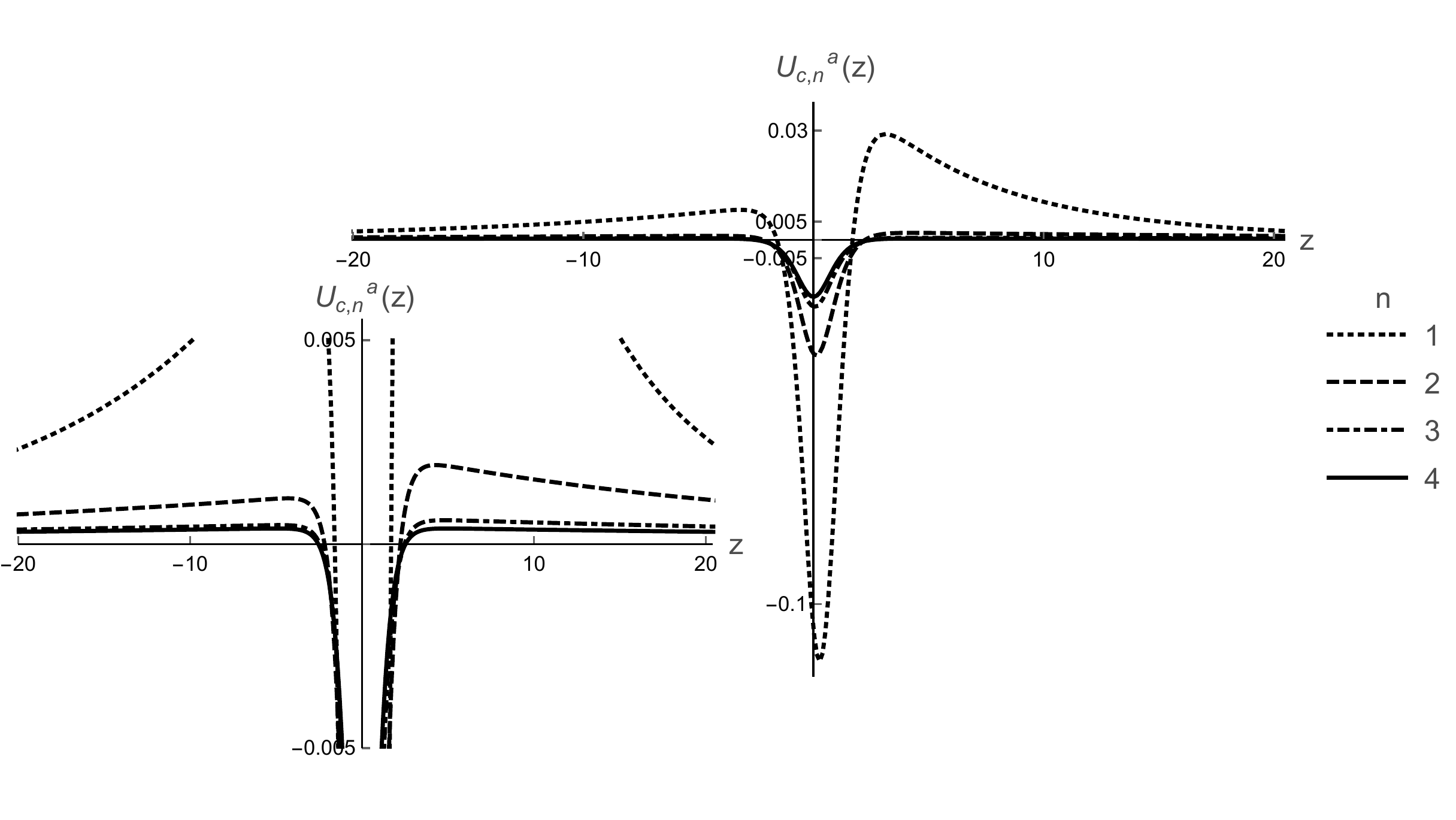}}
\caption{The stability potential (\ref{sp}) for $a=8$ and $n=1,2,3$ and 4}\label{fig14}
\end{figure}
%%%%%%%%%%%%%%%%%%%%%%%%%%%%%%%%%%%%%%%%%%%%%%

%%%%%%%%%%%%%%%%%%%%%%%%%%%%%%%%%%%%
\subsection{Metric fluctuations}
Let's study the stability of the gravitational sector in this section. The standard procedure here is to perform a variable redefinition $ dy^ 2 \rightarrow  e^{2A(z)} dz^2 $ in (\ref{metric}) and rewrite the conformally flat metric metric $\tilde{g}_{ab}=e^{2A(z)}\eta_{ab}$ with a linear pertubation as
\begin{equation}
ds^2=e^{2A(z)}\left(\eta_{ab}+h_{ab}\right)dx^a dx^b.
\end{equation}
In the  transverse-traceless  gauge we have $\partial_{\mu}h^{\mu\nu}=0$ and $h_{\mu}^{\mu}=0$. Moreover the conformal Einstein tensor takes the simple form ${\bar G}=-\frac12\partial_c\partial^c h_{ab}$, leading to linearized Einstein tensor
\begin{eqnarray}
\nonumber G_{ab}^{(1)}&=&-\frac{1}{2}\partial_c\partial^c h_{ab}+3\Bigl[\partial_a A\partial_b A-\partial_a \partial_b A+\\
&~&+\frac{1}{2}A'h'_{ab}+\bar{g}_{ab}\left(\partial_c \partial^c A+\partial_c A\partial^c A\right)\Bigr].
\end{eqnarray}
In this configurattion the $\mu\nu-$components of $G_{ab}^{(1)}$ are
\begin{equation}
G_{\mu\nu}^{(1)}=-\frac{1}{2}\partial_c\partial^c h_{\mu\nu}+\frac{3}{2}A'h'_{\mu\nu}-3\bar{g}_{\mu\nu}\left( A''+ A'^2\right).
\end{equation}
and the linearized energy-momentum tensor becomes
\begin{equation}
T_{\mu\nu}^{(1)}=-\frac{3}{2}\bar{g}_{\mu\nu}\left( A''+ A'^2\right)
\end{equation}
where the prime denotes the derivative in relation to variable $z$. Hence, we use linearized Einstein equations, $G_{\mu\nu}^{(1)}=2T_{\mu\nu}^{(1)}$, to find the equation for $h_{\mu\nu}$, wich is given by $-\partial_c\partial^c h_{\mu\nu}+3A'h'_{\mu\nu}=0$. The final step is to perform the redefinition $H_{\mu\nu}=e^{-ipx}e^{3A/2}h_{\mu\nu}$ and rewrite the equation for $h_{\mu\nu}$ in terms of $H_{\mu\nu}$, wich is 
\begin{equation}\label{eqH}
\left(\partial_z+\frac{3}{2}A'\right)\left(-\partial_z+\frac{3}{2}A'\right)H_{\mu\nu}=p^2 H_{\mu\nu}.
\end{equation}
Equation (\ref{eqH}) represents a Schr\"odinger equation for a  Supersymmetric Quantum Mechanics problem  with the stability potential given by
\be\label{sp}
U(z)=\frac32 A^{\prime\prime}+ \frac94 A^{\prime 2}.
\ee
wich is  depicted in (\ref{fig14}). The factorized form of (\ref{eqH}) is $S^{+}S^{-}\psi=p^2\psi$, with $S^{\pm}=\left(\pm\partial_z+3A'/2\right)$, wich indicates that there are no negative  gravitons modes, since the Hermitian operator $S^{+}S^{-}$ is non-negative. It ensures system stability.
%%%%%%%%%%%%%%%%%%%%%%%%%%%%%%
%%%%%%%%%%%%%%%%%%%%%%%%%%%%%%%%%%%%%%%%%%%%%%
\section{Ending comments}
In this work, new models of asymmetric kinks were constructed for Field Theory, which in turn stimulated the investigation of new models of asymmetric braneworlds. To construct such models, we use the Deformation Method developed in \cite{blm}. Deforming a previously studied model, we arrive at two new classes of solutions written in terms of the Chebyshev Polynomials and controlled by two parameters, where one specifies each model of the presented classes and the other gives the number of topological defects present in each model.

Both topological sectors  of (\ref{potc}) and (\ref{modelu}) behaves qualitatively equals, but the  model (\ref{modelu}) has a peculiar characteristic that associates it with systems that exhibit some kind of asymptotic freedom.% An interesting fact is that some of the non-topological sectors of the (\ref{modelu}) model can be described as trajectories of particles that evolve from a rest state to a stationary state, and vice versa. 

Finally, we analyze the braneworld that can be constructed from the class (\ref{potc}). We obtain a tower of asymmetric branes modeled by two parameters, one that specifies the amount of branes that can be constructed for each model and another one that indicates which hierarchy they should be ordered in. We also obtained the expressions for the energy density and presented the stability of the model. 
%%%%%%%%%%%%%%%%%%%%%%%%%%%%%%%%%%%%%%%%%%%%%%%%%
%%%%%%%%%%%%%%%%%%%%%%%%%%%%%%%%%%%%%%%%%%%%%%%%%
\acknowledgements{We thank Dionisio Bazeia for comments and discussions. We also thank the Brazilian agencies CAPES and CNPq for financial support.}
%%%%%%%%%%%%%%%%%%%%%%%%%%%%%%%%%%%%%%%%%%%%%%%%%
%%%%%%%%%%%%%%%%%%%%%%%%%%%%%%%%%%%%%%%%%%%%%%%%%

%%%%%%%%%%%%%%%%%%%%%%%%%%%%%%%%%%%%%%%%%%%%%%%%%%
%%%%%%%%%%%%%%%%%%%%%%%%%%%%%%%%%%%%%%%%%%%%%%%%%%

\end{document}